\documentclass[a4paper,10pt]{article}
\pdfoutput=1 

\usepackage{jheppub} 

\usepackage[T1]{fontenc} 
\usepackage{graphicx,color}
  \usepackage{bm,booktabs}
   \usepackage{amsmath}
    \usepackage{amssymb}
     \usepackage{pifont}
      \usepackage{simplewick}
\usepackage{tikz}
\usepackage[most]{tcolorbox}
\usepackage{rotating}
\usepackage[normalem]{ulem}

\newcommand{\ot}{\leftarrow}

\renewcommand{\(}{\left(}
\renewcommand{\)}{\right)}
\renewcommand{\[}{\left[}
\renewcommand{\]}{\right]}

\renewcommand{\vec}[1]{\bm{#1}}

\newcommand{\specialcellcenter}[2][c]{\begin{tabular}[#1]{@{}c@{}}#2\end{tabular}}

\begin{document}
\title{Extraction of unpolarized quark transverse momentum dependent parton distributions from Drell-Yan/Z-boson production}

\author[a]{Valerio Bertone,}
\affiliation[a]{Dipartimento di Fisica, Universit\`a di Pavia and INFN, Sezione di Pavia Via Bassi 6, I-27100 Pavia, Italy}
\emailAdd{valerio.bertone@cern.ch}
\author[b]{Ignazio Scimemi,}
\affiliation[b]{Departamento de F\' isica Te\'orica and IPARCOS, Universidad Complutense de Madrid,
Ciudad Universitaria,  28040 Madrid, Spain}
\emailAdd{ignazios@fis.ucm.es}
\author[c]{Alexey Vladimirov}
\affiliation[c]{Institut f\"ur Theoretische Physik,  Universit\"at Regensburg,
D-93040 Regensburg, Germany}
\emailAdd{alexey.vladimirov@ur.de}

\abstract{ 
We present the extraction of unpolarized quark transverse momentum dependent parton distribution functions (TMDPDFs) and the non-perturbative part of TMD evolution kernel from the global analysis of Drell-Yan and $Z$-boson production data. The analysis is performed at the next-to-next-to-leading order (NNLO) in perturbative QCD, using the $\zeta$-prescription. The estimation of the error-propagation from the experimental uncertainties to non-perturbative function is made by the replica method. The importance of the inclusion of the precise LHC data and its influence on the determination of non-perturbative functions is discussed.
}
\maketitle

\section{Introduction}

The description of the hadron structure is one of the major challenges for the comprehension of strong interactions. Transverse momentum dependent parton distribution functions (TMDPDFs) depict parton momenta in  3-dimensions and provide more detailed information on hadrons than the one-dimensional collinear parton distribution functions (PDFs). In this work, we present the extraction of unpolarized TMDPDF and non-perturbative part of TMD evolution from the fit of Drell-Yan and Z-boson production measurements.

At hadron colliders, in the regime of the small transverse momentum of the produced vector/scalar-boson, the cross-section is factorizable in terms of universal TMDPDFs \cite{Collins:2011zzd,GarciaEchevarria:2011rb,Vladimirov:2017ksc}. The phenomenological analysis of Drell-Yan and Z-boson production processes (we refer to them as Drell-Yan (DY) processes, for simplicity) within the TMD factorization has a long history, see e.g. ref.~\cite{Landry:1999an,Qiu:2000hf,Landry:2002ix,Watt:2003vf,Mantry:2010bi,Becher:2010tm,Su:2014wpa,DAlesio:2014mrz,Bacchetta:2017gcc,Scimemi:2017etj}. However, many of these works  have been produced before a rigorous formulation of the TMD factorization and TMD evolution and for that reason are outdated.  These articles differ, among the others, in the phenomenological construction of the factorized cross-section (which is relevant for the theoretical precision that can be achieved), the composition of perturbative and non-perturbative contributions and the inspected data sets.  Also, the majority of the fits included in this list operates only at perturbative leading order (LO)  and do not include the highly precise measurements made at LHC. In the present work, we aim to cover this gap and to obtain precise values of the TMDPDFs and of the non-perturbative part of the TMD evolution consistently with modern theory and data.

Over the past few years the theory of TMD factorization has  developed consistently. In particular, nowadays its perturbative structure  is completely understood, which is confirmed by multiple next-to-next-to-leading order (NNLO) perturbative calculations \cite{Catani:2012qa,Catani:2013tia,Gehrmann:2014yya,Echevarria:2015byo,Echevarria:2015usa,Echevarria:2016scs,Vladimirov:2016dll,Li:2016ctv,Gutierrez-Reyes:2018iod}. Also there was a critical progress in the understanding of the structure of TMD evolution \cite{Collins:2011zzd,Chiu:2012ir,Echevarria:2012js,Scimemi:2016ffw,Vladimirov:2017ksc}, and the relation between different components of TMD scaling \cite{Collins:2016hqq,Scimemi:2018xaf}. For a recent review of the  state-of-the-art, we refer to \cite{Scimemi:2019mlf, Angeles-Martinez:2015sea}. The present extraction is founded on these theory achievements and uses the highest  perturbative input available nowadays, that is, the complete NNLO (two-loop coefficient functions together with three-loop evolution).
 
The extraction of the TMDPDF requires an articulated consideration of the scale settings, which is performed here using the $\zeta$-prescription. Since the approach is novel in the TMD factorization studies, we explain its origin and importance in a few words, and we refer to the original paper~\cite{Scimemi:2018xaf} for the details. The $\zeta$-prescription consists of a particular choice of renormalization and rapidity evolution scales for TMD distributions.
The double scale dependence is characteristic of the TMD distinctions, and it can be traced in perturbative calculation due to the different origin of divergences. The presence of two scales results in a non-elementary problem of the scale-fixation choice for TMD distributions. Within $\zeta$-prescription the TMD evolution is made effectively one-dimensional, which allows selecting the best values for the scale parameters (this choice is known as an optimal TMD distribution) that guarantee the perturbative stability. As a major outcome, the $\zeta$-prescription consistently separates the non-perturbative part of the evolution kernel from the non-perturbative parton distribution. For this reason, the values of the non-perturbative evolution extracted in this work are universal and can be used directly in other applications, e.g., the analysis of polarized TMD distributions \cite{Anselmino:2016uie, Anselmino:2018psi} or jet productions \cite{Kang:2017btw, Gutierrez-Reyes:2018qez}.

Beyond the modern state-of-the-art implementation of TMD factorization, here we reconsider the extraction of TMDPDF including a larger set of experimental data and we provide a solid statistical analysis of error-propagation. Comparing this fit with the most recent and complete extractions made in refs.~\cite{Bacchetta:2017gcc,Scimemi:2017etj}, the number of analyzed data points is significantly bigger (457 points against  293 in \cite{Bacchetta:2017gcc} and 309 in \cite{Scimemi:2017etj}, which is the biggest amount of DY data ever considered, to our knowledge). This number of data has been achieved by including the results from PHENIX \cite{Aidala:2018ajl}, E772 \cite{McGaughey:1994dx} experiments, differential rapidity bins from ATLAS \cite{Aad:2015auj} and the measurement of the Drell-Yan cross-section in the muon channel at D0 \cite{Abazov:2010kn}. These data points are included in the analysis of TMD cross-section for the first time\footnote{Let us mention, that the LHC data has also been analyzed in the resummation approach \cite{Bozzi:2010xn,Catani:2015vma, Bizon:2018foh} with the same level of perturbative input. However, the resummation approach should not be confused with the TMD factorization, although they have several common points. The resummation approach is founded on  collinear factorization, and it has theoretically no access to a non-perturbatively generated transverse momentum. For that reason, the resummation approach is only applicable at high-energy and at larger values of $q_T$.}.  For the determination of the extraction uncertainties we apply the replica method~\cite{Ball:2008by,Ball:2014uwa,Ball:2017nwa,Bertone:2017tyb}, routinely used for the extraction of collinear PDFs. We have found that the inclusion of the LHC data essentially reduces the uncertainty band for non-perturbative functions. Nonetheless, the available data leave uncovered a large portion of the energy/momentum phase space that should be filled by experiments in the future. 

As a result, we obtain a consistent and complete picture of the unpolarized TMDPDFs and their evolution kernel supporting it with a well established statistical treatment. We think that such screening is fundamental to provide clear indications to experimentalists and theorists about the validity of the TMD factorization theorem, and it represents a notable improvement in the understanding of transverse momentum structure of a hadron. The results of this work are available as a part of \texttt{artemide}-package for TMD phenomenology \cite{web}. The library contains the routines for the evaluation of TMDPDFs and their evolution (mean values and distribution of replicas) and the routines for the evaluation of the related cross-section.

The paper is organized as follows. In sec.~\ref{sec:DY} we review the TMD factorization and the necessary elements of the theory, such as TMD evolution and $\zeta$-prescription in sec.~\ref{sec:TMDev}, fundamental requirements on the model building and collinear matching in sec.~\ref{sec:model-requirements}. We formulate the non-perturbative models for rapidity anomalous dimension in sec.~\ref{sec:RADmodel}, and for TMDPDF in sec.~\ref{sec:TMDmodel}. The selection of the data set is discussed in sec.~\ref{sec:data}, while the details of the statistical analysis can be found in sec.~\ref{sec:stat} and in the appendices. Finally, we present the results in sec.~\ref{sec:results}. In particular, the quality of the fit is discussed in sec.~\ref{sec:agree} and the extracted non-perturbative functions are discussed in sec.~\ref{sec:values}.

\section{Drell-Yan cross section in TMD factorization}
\label{sec:DY}

The leading term of the TMD-factorized cross section for the DY process ($h_1+h_2\to Z/\gamma^*(\to ll')+X$) has the following structure~\cite{Tangerman:1994eh,GarciaEchevarria:2011rb,Collins:2011zzd}
\begin{eqnarray}\label{def:xSec}
\frac{d\sigma}{dQ^2 dy dq_T^2}=\sigma_0\sum_{f_1,f_2}H_{f_1f_2}(Q,\mu)\int \frac{d^2\vec b}{4\pi} e^{i(\vec b\cdot \vec q_T)}F_{f_1\ot h_1}(x_1,\vec b;\mu,\zeta_1)F_{f_2\ot h_2}(x_2,\vec b;\mu,\zeta_2),
\end{eqnarray}
where $Q^2= (l+l')^2$, $\vec q_T$ and $y$ are transverse component and rapidity of the lepton pair momentum with respect to collision axis, and the variables $x_{1,2}$ are defined as
\begin{eqnarray}\label{def:x12}
x_{1,2}=\frac{\sqrt{Q^2+\vec q_T^2}}{\sqrt{s}}e^{\pm y}\,.
\end{eqnarray}
The function $F_{f\to h}$ is the unpolarized TMDPDF\footnote{Traditionally, the unpolarized TMDPDF is denoted as $f_1(x,\vec b)$. Here, we use the notation $F(x,\vec b)$ in order to avoid any confusion with the collinear function $f(x,\mu)$ and non-perturbative ansatz $f_{NP}(x,\vec b)$ introduced in the following.} of the parton flavor $f$ in hadron $h$ in impact parameter space $\vec b$. The function $H$ is the hard-scattering coefficient function and $\sigma_0$ is a kinematic factor. For a more detailed definition, we refer the reader to ref.~\cite{Scimemi:2017etj,Becher:2010tm,Becher:2011xn}. The factorization formula in eq.~(\ref{def:xSec}) is accurate to leading power in $\vec q_T^2/Q^2$, while power-suppressed corrections are presently unknown (see ref.~\cite{Balitsky:2017gis,Ebert:2018gsn} for recent developments).

The scales $\mu$ and $\zeta_{1,2}$ are the renormalization and rapidity scales, respectively~\cite{Collins:2011zzd,Vladimirov:2017ksc,Echevarria:2012js,Chiu:2012ir}. In order to minimize the logarithms in hard coefficient function $H$, we set the renormalization scale $\mu$ equal to the hard scale $Q$. Moreover, the rapidity scales must obey the relation $\zeta_1\zeta_2=Q^4$: we make the symmetric choice $\zeta_1=\zeta_2=Q^2$.

In the following of this section, we briefly review the relevant ingredients of eq.~(\ref{def:xSec}), discussing the TMD evolution and the separation between perturbative and non-perturbative components. Then we describe the models used to parametrize the non-perturbative input. Finally, we give the final expression for the cross section and discuss the perturbative input used for the fits.

\subsection{TMD evolution}
\label{sec:TMDev}

In order to consistently combine the perturbative and non-perturbative parts of the TMD factorization formula (\ref{def:xSec}), and to separate the matching and evolution effects within TMDPDFs, we use the $\zeta$-prescription. It is based on the notion of double-scale evolution, and consists in a special definition evolution scale. We refer to ref.~\cite{Scimemi:2018xaf} for a detailed description of the double-scale evolution and its properties. In this section, we present minimal introduction to $\zeta$-prescription and formulas that are used in the fit.

The TMD evolution in the $(\mu,\zeta)$-plane is governed by the pair of differential equations whose kernels define a bi-dimensional scalar potential. The logarithm of the TMD evolution factor $R$ is given by the difference between potentials at different points of $(\mu,\zeta)$-plane, and for that reason, TMD distribution evaluated on two points with the same value of potentials are equal. Within the $\zeta$-prescription, a TMD distribution is defined by an equipotential line, instead of the scales $(\mu,\zeta)$, and it evolution is given by a transition between equipotential lines. 

The line that goes through the saddle point of the potential is \textit{special}, since it is a uniquely and non-perturbatively defined, and spans the whole range in $\mu$ and $\zeta$. This line provides a natural starting point for the definition of the non-perturbative component of TMD distributions.  Given $\zeta=\zeta_\mu(\vec b)$ belonging to the special line\footnote{This approach, dubbed $\zeta$-prescription, has been proposed in ref.~\cite{Scimemi:2017etj}. A comprehensive discussion on this prescription and the definition of the \textit{optimal} TMD can be found in ref.~\cite{Scimemi:2018xaf}.}, we define the \textit{optimal} TMD distribution as
\begin{eqnarray}
F_{f\ot h}(x,\vec b;\mu,\zeta_\mu(\vec b))=F_{f\ot h}(x,\vec b)\quad \mu\in\mbox{ special line},
\end{eqnarray}
where in the r.h.s. we have emphasized its ``naive scale-independence''.
The evolution of the optimal TMD distribution to a generic set of scales $(\mu,\zeta)$ is then simply given by
\begin{eqnarray}\label{th:evolution}
F_{f\ot h}(x,\vec b;\mu,\zeta)=R^f[\vec b; (\mu,\zeta)\to(\mu_0,\zeta_{\mu_0}(\vec b))]F_{f\ot h}(x,\vec b),
\end{eqnarray}
where $R^f$ is the TMD evolution factor whose expression is
\begin{eqnarray}
\label{eq:R}
R^f[\vec b;(\mu_1,\zeta_1)\to(\mu_2,\zeta_2)]=\exp\[\int_P \(\frac{\gamma_F^f(\mu,\zeta)}{2}\frac{d\mu^2}{\mu^2}-\mathcal{D}^f(\mu,\vec b)\frac{d\zeta}{\zeta}\)\].
\end{eqnarray}
Note that the r.h.s. of eq.~(\ref{th:evolution}) is effectively independent on $\mu_0$. The anomalous dimension $\gamma_F$ and rapidity anomalous dimension $\mathcal{D}$ are universal for all TMD distributions and their perturbative expressions are currently known up to three-loop~\cite{Moch:2005tm,Baikov:2009bg,Vladimirov:2016dll,Li:2016ctv}. Importantly, the rapidity anomalous dimension has a non-perturbative component that is usually extracted from data along with the non-perturbative component of TMD distributions.

The integration path $P$ in eq.~(\ref{eq:R}), that connects the points $(\mu_1,\zeta_1)$ and $(\mu_2,\zeta_2)$ in the evolution plane, is in principle arbitrary.  In practice, the evolution factor $R^f$ is independent on the path $P$ only if all terms in the perturbation expansion of the anomalous dimensions are included. This property is violated by the truncation of perturbative expansion. However, one can define a scheme for the evolution that preserves the conservativeness of the potential. Clearly, the difference between schemes tends to vanish as more and more terms are included in the perturbative expansions. In this work, we use the so-called improved-$\gamma$ scheme defined in ref.~\cite{Scimemi:2018xaf}. For the numerical implementation of the evolution factor we use the simplest possible path, \textit{i.e.} a straight line that connects $\zeta$ to $\zeta_\mu(\vec b)$ at fixed $\mu$. By doing this, the evolution factor takes the form
\begin{eqnarray}\label{th:evolution_our}
R^f[\vec b;(\mu,\zeta)\to (\mu,\zeta_\mu(\vec b))]&=&R^f[\vec b;(\mu,\zeta)]=\(\frac{\zeta}{\zeta_\mu(\vec b)}\)^{-\mathcal{D}^f(\mu,\vec b)}.
\end{eqnarray}
Remarkably, this expression does not involve any integration. This entails a great simplification of the numerical implementation of the TMD evolution.

\subsection{General requirements for the TMD distributions}
\label{sec:model-requirements}

The non-perturbative parts of the TMDPDF $F$ and the rapidity anomalous dimension are to be extracted from data. However, a number of theoretically justified constraints can be enforced.
\begin{itemize}
\item For $\vec b \to 0$, the non-perturbative component of both TMD distributions and rapidity anomalous dimension is expected to be suppressed. In particular, in this regime TMDPDF can be computed as
\begin{eqnarray}\label{model:smallb}
\vec b \to 0, \qquad F_{f\to h}(x,\vec b)=\sum_{f'}\int_x^1\frac{dy}{y}C_{f\ot f'}\(\frac{x}{y},\ln\(\vec b^2 \mu^2\)\)f_{f'\ot h}(y,\mu),
\end{eqnarray}
where $f_{f\ot h}$ is the collinear PDF for the parton flavor $f$. The coefficient functions $C$ are currently known up to two-loop order~\cite{Echevarria:2015usa,Echevarria:2016scs}. 

\item The leading power correction to the small-$\vec b$ is of order $\vec b^2$. This follows from the operator product expansion and it has been confirmed by the explicit evaluation of the renormalon contributions~\cite{Scimemi:2016ffw}. In general, power corrections to the small-$\vec b$ must scale as $\vec b^{2n}$, \textit{i.e.} only even powers of $\vec b$ are allowed in the Taylor expansion around $\vec b=0$.

\item The asymptotic for $\vec b\to \infty$ is mostly unknown. A reasonable restriction is that both TMDs and evolution factor should tend to zero in this limit. However, the decay law is unknown. Typical choices are a gaussian or an exponential falloff.
\end{itemize}
These restrictions significantly constrain the behavior of the non-perturbative components, particularly at small $\vec b$. At large $\vec b$, instead, theoretical constraints are milder. Based of these considerations, in the following we propose models for the rapidity anomalous dimension and the intrinsic part of TMDPDFs. 

\subsection{Model for rapidity anomalous dimension}
\label{sec:RADmodel}

The non-perturbative rapidity anomalous dimension $\mathcal{D}^f$ is modeled by the following function
\begin{eqnarray}\label{model:rad}
\mathcal{D}^f(\mu,\vec b)=\mathcal{D}^f_{\text{res}}\(\mu,b^*(\vec b)\)+g(\vec b),
\end{eqnarray}
where $\mathcal{D}^f_\text{res}$ is the resummed perturbative part of $\mathcal{D}^f$, $g$ is an even function of $\vec b$ vanishing as $\vec b\to 0$, and
\begin{eqnarray}\label{eq:bstar}
b^*(\vec b)=\sqrt{\frac{\vec b^2 B_{\text{NP}}^2}{\vec b^2+B_{\text{NP}}^2}}\,.
\end{eqnarray}
The resummed anomalous dimension $\mathcal{D}^f_{\text{res}}$ can be expanded as
\begin{eqnarray}\label{model:d_resum}
\mathcal{D}^f_{\text{res}}\(\mu,\vec b\)=\sum_{n=0}^\infty a_s^n(\mu) d^f_n(X),
\end{eqnarray}
where $X=\beta_0 a_s(\mu)\ln(\mu^2\vec b^2 e^{2\gamma_E}/4)$, with $a_s=g^2/(4\pi)^2$. The leading term reads
\begin{eqnarray}\label{model:d0}
d^f_0(X)=-\frac{\Gamma_0^f}{2\beta_0}\ln(1-X),
\end{eqnarray}
where $\beta_0$ is the leading-order (LO) coefficient of the expansion of the QCD $\beta$-function and $\Gamma_0^f$ is LO cusp anomalous dimension  ($\beta_0=(11 C_A-2 N_f)/3$ and $\Gamma_0=4 C_F$, respectively). For our studies we have used eq.~(\ref{model:d_resum}) at NNLO (\textit{i.e.} up to $d_2^f$). The NNLO expression incorporates the three-loop anomalous dimension and can be found in ref.~\cite{Echevarria:2012pw,Scimemi:2018xaf}. 

Due the definition of $d^f_0$ in eq.~(\ref{model:d0}), the resummed rapidity anomalous dimension is singular at $X=1$. Roughly, it corresponds to $\vec b^2\sim 4e^{-2\gamma_E}/\Lambda_{QCD}^2\simeq (4.5\,\text{GeV}^{-1})^2$, which is deep in the non-pertrubative region of $\vec b$. In order to avoid the singularity, we replace $\vec b$ with $b^*$ defined in eq.~(\ref{eq:bstar}) in the resummed part of the anomalous dimension. Since $b^*$ never exceeds $B_{\text{NP}}$, the value of $\mathcal{D}_{\text{res}}$ approach $\mathcal{D}_{\text{res}}(\mu,B_{\text{NP}})$ at large $\vec b$. The function $g(\vec b)$ in eq.~(\ref{model:rad}) represents the non-perturbative contribution to the anomalous dimension. Based on general considerations, the Taylor expansion around $\vec b=0$ of this function contains only even powers of $\vec b$, starting from $\vec b^2$. Therefore, generally, the model (\ref{model:rad}) satisfies all requirements listed in sec.\ref{sec:model-requirements}.

In our research we have tested different models for $g(\vec b)$. We have found that, the current data do not allow for an accurate extraction of the function $g$ at large-$\vec b$. Practically, only the leading term $\sim\vec b^2$ could be rigorously fixed, and it should be small enough, so it does not affect the small-$\vec b$ part (that is fixed by perturbation theory). Finally, we have adopted a simple one-parameter exponential model
\begin{eqnarray}\label{model:exp}
g(\vec b)=c_0 \vec b b^*(\vec b).
\end{eqnarray}
At small $\vec b$ this model behaves as  $g(\vec b)\sim \lambda_0 \vec b^2$, whereas, at large $\vec b$, instead, it behaves as $g(\vec b)\sim \lambda_0 \vec b B_{\text{NP}}$. The other candidate for the final model of non-perturbative evolution was a more traditional Gaussian model, $g(\vec b)\sim c_0 \vec b^2$ (see ref.~\cite{Collins:2014jpa} for a recent review). However, since exponential and Gaussian models provide a similar description of the experimental data, we find preferable to use the exponential model in eq.~(\ref{model:exp}). The reason is that it appears to extend the validity of the perturbative series to higher values of $\vec b$.

\subsection{Model for TMDPDF}
\label{sec:TMDmodel}

In our fits, the model that parametrizes the intrinsic non-perturbative component of the TMDPDFs is implemented by means of the following general form
\begin{eqnarray}\label{model:TMDPDF}
F_{f\to h}(x,\vec b)=f_{\rm NP}(x,\vec b)\sum_{f'}\int_x^1\frac{dy}{y}C_{f\ot f'}\(\frac{x}{y},\ln\(\vec b^2 \mu^2\)\)f_{f'\ot h}(y,\mu),
\end{eqnarray}
where $f_{\rm NP}$ is a function to be fitted to data.
Eq.~(\ref{model:TMDPDF}) is not the most general ansatz that satisfies the requirements discussed in the previous section. In particular, $f_{\rm NP}$ may depend on the flavor and also on the convolution variable $y$, but we have found that this ansatz is sufficient to describe the data at the current level of precision.

The factorization scale $\mu$ in the r.h.s. of eq.~(\ref{model:TMDPDF}) is chosen to be
\begin{eqnarray}
\mu=\frac{2e^{-\gamma_E}}{|\vec b|}+2\,\text{GeV}\,.
\end{eqnarray}
This choice allows the impact parameter $|\vec b|$ not to reach the Landau pole. In any case it was  found  that the dependence on the exact value of the scale is not very large~\cite{Scimemi:2018xaf}. Concerning the input collinear PDFs $f_{f'\ot h}$, we have tried different publicly available sets and found that there is a marked dependence on the particular choice. It implies that the TMD physics is sensitive to the $x$-dependence at small-$\vec b$, which is totally dictated by choice of PDF set by contraction of our model (\ref{model:TMDPDF}). We leave a detailed study of this dependence for a future publication. For the current fit, we have used the central replica of the NNPDF3.1 NNLO set~\cite{Ball:2017nwa} through the LHAPDF library~\cite{Buckley:2014ana}. This set provides the best description of the data.
The LHAPDF library also provides the strong running coupling $\alpha_s$ consistently with the PDF set\footnote{
The transition between perturbative/non-perturbative regimes in model (\ref{model:rad}) for $\mathcal{D}$ takes a place at $\alpha_s(\sim 1 \text{GeV})$. Therefore, it is significantly influenced by a particular realization of running $\alpha$ at small values of $\mu$. In this way, out choice of PDF set indirectly affects non-pertrubative part of the evolution.}. The flavor number $N_f$ is so consistently and automatically fixed  at the correct  scale through $\alpha_s$ and ultimately the PDF sets.

The shape of the function $f_{\rm NP}$ significantly influences the value of the cross section. Therefore, in order to avoid possible parametric biases, it should be chosen to be as flexible as possible taking into account  the following theoretical constraints. First, $f_{\rm NP}$ has to be such that $\lim_{\vec b\to0}f(x,\vec b)=1$. Second, it should be an even function of $\vec b$, \textit{i.e.} the Taylor expansion around $\vec b=0$ should only contain even powers of $\vec b$. We have found that a suitable parametrization of $f_{\rm NP}$ has the form
\begin{eqnarray}
\label{eq:fnp}
f_{\rm NP}(x,\vec b)=\exp\(-\frac{r_1(x,\vec b)}{r_2(x,\vec b)}\),
\end{eqnarray}
where at small $\vec b$ $r_1(x,\vec b)\sim r_1(x,0) \vec b^2+...$ and $r_2(x,\vec b)\sim 1+...$. The function $r_1$ gives dominant behavior at small-$\vec b$, whereas the function $r_2$ controls the large-$\vec b$ region. The Pad\'e-like form of the exponent guaranties that the higher powers of $\vec b$ do not give a large contribution. Therefore the functions $r_1$ and $r_2$ can be expanded around $\vec b = 0$ and truncated after the first few terms. We have performed numerous tests and found that the current data do not resolve the higher modes of the $x$-dependence, and thus the functions $r_{1}$ and $r_2$ can be simple polynomials in $x$.
Specifically, we use the following model
\begin{eqnarray}\label{model:our_fNP}
f_{\rm NP}(x,\vec b)=\exp\(-\frac{(\lambda_1(1-x)+\lambda_2 x+\lambda_3 x(1-x))\vec b^2}{\sqrt{1+\lambda_4 x^{\lambda_5} \vec b^2}}\),
\end{eqnarray}
where $\lambda_{1,..,5}>0$. This parametrization, with five free parameters, is able to accommodate a range of different behaviors, such as the exponential and the Gaussian one, with some degree of redundancy.  Specifically, we have found that the number of free parameters can be reduced to three or four without a significant deterioration in the description of the data.

\subsection{Summary on theory input}
\label{sec:theory_sum}

The final formula to compare to the DY experimental data is 
\begin{align}\label{def:xSec_p}
\frac{d\sigma}{dQ^2 dy dq_T^2}=\sigma_0\sum_{f_1,f_2}H_{f_1f_2}(Q,Q)\int \frac{d^2\vec b}{4\pi} e^{i\vec b\cdot \vec q_T}\{R[\vec b;(Q,Q^2)]\}^2 F_{f_1\ot h_1}(x_1,\vec b)F_{f_2\ot h_2}(x_2,\vec b).
\end{align}
The explicit form of the  TMDPDFs $F$ is given in eq.~(\ref{model:TMDPDF}) with the non-perturbative input given in eq.~(\ref{model:our_fNP}). The expression for the TMD evolution factor is given in eq.~(\ref{th:evolution_our}). The model used for $\mathcal{D}$ anomalous dimension is given in eq.~(\ref{model:rad}) with the non-perturbative input given in eq.~(\ref{model:exp}). In conclusion, in our fit there is a total of seven free parameters (two for the evolution and five for the TMDPDFs). The summary of the  perturbative input used for the computation of the observables is presented in Table~\ref{tab:pert}.
\begin{table}[h]
\begin{center}
\begin{tabular}[h]{|c||c|c||c|c|c||c|c|}\hline
Function & $H$ &  $C_{f\ot f'}$ & $\Gamma_{\text{cusp}}$ & $\mathcal{D}$ & $\gamma_F$ & $\alpha_s$ running & PDF evolution
\\\hline 
Order & $\alpha_s^2$ & $\alpha_s^2$ & $\alpha_s^3$ & \specialcellcenter{$\alpha_s^2$\\ resummed} & $\alpha_s^3$ & \multicolumn{2}{|c||}{\specialcellcenter{NNLO provided by \\ NNPDF3.1~\cite{Ball:2017nwa}} }
\\\hline
\end{tabular}
\end{center}
\caption{\label{tab:pert} Summary of perturbative orders used in the fit for each part of the cross section.}
\end{table}

\section{Data selection}
\label{sec:data}

The TMD factorization of the cross section is valid only in small transverse-momentum ($q_T$) regime. Therefore, we need to impose  a cut on the experimental data set that limits the kinematics of the data points to this region. In our fit we have selected the data according to the following rule: given a data point $p\pm \sigma$, with $p$ being the central value and $\sigma$ its uncorrelated relative uncertainty, corresponding to some values of $q_T$ and $Q$ (which are taken to be the center of the bin), we include it in the fit only if
\begin{eqnarray}\label{cuts}
\delta\equiv \frac{q_T}{Q} <0.1,\qquad \text{or}\qquad \delta<0.25\quad \text{if}\quad \delta^2<\sigma.
\end{eqnarray}
These conditions are chosen for the following reasons. In ref.~\cite{Scimemi:2017etj} it has been demonstrated that, within the experimental accuracy of the data set included in the fit, TMD factorization is valid in the range $\delta<(0.1-0.25)$ . At higher values of $\delta$, power corrections to TMD factorization, that scale as $q_T^2/Q^2=\delta^2$, should be taken into account. Specifically, in the TMD framework, these corrections can be regarded as a theoretical uncertainty. Based on this consideration, if the (uncorrelated) experimental uncertainty of a given data point is smaller than the theoretical uncertainty associated to the expected size of power corrections, we drop this point from the fit. This is the origin of the second condition in eq.~(\ref{cuts}). This data selection is particularly conservative because it drops points that could potentially be described by TMD factorization (see e.g. ref.~\cite{Bacchetta:2017gcc} where less conservative cuts are used). However, this choice guarantees that we operate well within the range of validity TMD factorization.

\begin{table}[t]
\begin{center}
\small
\begin{tabular}{|c||c|c|c|c|c|c|}
\hline
Experiment & ref. 
&$\sqrt{s}$ [GeV]& $Q$ [GeV] & $y$/$x_F$ & \specialcellcenter{fiducial\\region}  & \specialcellcenter{$N_{\rm pt}$\\after cuts}
\\\hline\hline 
E288 (200) & \cite{Ito:1980ev} 
& 19.4 & \specialcellcenter{4 - 9 in\\ 1~GeV bins$^*$} & $0.1<x_F<0.7$  & -& 43
\\\hline
E288 (300) & \cite{Ito:1980ev} 
& 23.8 & \specialcellcenter{4 - 12 in \\ 1~GeV bins$^*$} & $-0.09<x_F<0.51$ & - & 53
\\\hline
E288 (400) & \cite{Ito:1980ev} 
& 27.4 & \specialcellcenter{5 - 14 in \\ 1~GeV bins$^*$} & $-0.27<x_F<0.33$ & - & 76
\\\hline\hline
E605 & \cite{Moreno:1990sf} 
& 38.8 & \specialcellcenter{7 - 18 in \\ 5 bins$^*$} & $-0.1<x_F<0.2$ & & - 53
\\\hline\hline
E772 & \cite{McGaughey:1994dx} 
& 38.8 & \specialcellcenter{5 - 15 in \\ 8 bins$^*$} & $0.1<x_F<0.3$ & - & 35
\\\hline\hline
PHENIX & \cite{Aidala:2018ajl} 
& 200 & 4.8 - 8.2 & $1.2<y<2.2$ & - & 3
\\\hline\hline
CDF (run1) & \cite{Affolder:1999jh} 
& 1800 & 66 - 116 & - & - & 33
\\\hline
CDF (run2) & \cite{Aaltonen:2012fi} 
& 1960 & 66 - 116 & - & - & 39
\\\hline\hline
D0 (run1) & \cite{Abbott:1999wk} 
& 1800 & 75 - 105 & - & - & 16
\\\hline
D0 (run2) & \cite{Abazov:2007ac} 
& 1960 & 70 - 110 & - & - & 8
\\\hline
D0 (run2) & \cite{Abazov:2010kn} 
& 1960 & 65 - 115 & $|y|<1.7$ & \specialcellcenter{$p_T>15$ GeV \\ $|\eta|<1.7$} & 8
\\\hline\hline
ATLAS (7TeV) & \cite{Aad:2014xaa} 
& 7000 & 66 - 116 & \specialcellcenter{$|y|<1$\\ $1<|y|<2$ \\ $2<|y|<2.4$} & \specialcellcenter{$p_T>20$ GeV \\ $|\eta|<2.4$} & 15
\\\hline
ATLAS (8TeV) & \cite{Aad:2015auj} 
& 8000 & 66 - 116 & \specialcellcenter{$|y|<2.4$\\ in 6 bins} & \specialcellcenter{$p_T>20$ GeV \\ $|\eta|<2.4$} & 30
\\\hline
ATLAS (8TeV) & \cite{Aad:2015auj} 
& 8000 & 46 - 66 & $|y|<2.4$ & \specialcellcenter{$p_T>20$ GeV \\ $|\eta|<2.4$} & 3
\\\hline
ATLAS (8TeV) & \cite{Aad:2015auj} 
& 8000 & 116 - 150 & $|y|<2.4$ & \specialcellcenter{$p_T>20$ GeV \\ $|\eta|<2.4$} & 7
\\\hline\hline
CMS (7TeV) & \cite{Chatrchyan:2011wt} 
& 7000 & 60 - 120 & $|y|<2.1$ & \specialcellcenter{$p_T>20$ GeV \\ $|\eta|<2.1$} & 8
\\\hline
CMS (8TeV) & \cite{Khachatryan:2016nbe} 
& 8000 & 60 - 120 & $|y|<2.1$ & \specialcellcenter{$p_T>20$ GeV \\ $|\eta|<2.1$} & 8
\\\hline\hline
LHCb (7TeV) & \cite{Aaij:2015gna} 
& 7000 & 60 - 120 & $2<y<4.5$ & \specialcellcenter{$p_T>20$ GeV \\ $2<\eta<4.5$} & 8
\\\hline
LHCb (8TeV) & \cite{Aaij:2015zlq} 
& 8000 & 60 - 120 & $2<y<4.5$ & \specialcellcenter{$p_T>20$ GeV \\ $2<\eta<4.5$} & 7
\\\hline
LHCb (13TeV) & \cite{Aaij:2016mgv} 
& 13000 & 60 - 120 & $2<y<4.5$ & \specialcellcenter{$p_T>20$ GeV \\ $2<\eta<4.5$} & 7
\\\hline\hline
Total & & 
& & & & 457
\\\hline
\end{tabular}
\par
*Bins with $9\lesssim Q \lesssim 11$ are omitted due to the $\Upsilon$ resonance.
\caption{Summary table for the data included in the fit.\label{tab:data}. For each data set we report: the reference publication, the centre-of-mass energy, the coverage in $Q$ and $y$ or $x_F$, possible cuts on the fiducial region, and the number of data points that survive the cut in eq.~(\ref{cuts}).}
\end{center}
\end{table}

Table~\ref{tab:data} reports a summary of the full data set included in our fit. Remarkably, after imposing the cut in eq.~(\ref{cuts}), the number of data points included in our fit is 457. Despite the conservative cut, this is the largest set of DY data considered so far within a TMD fit.  Our data set spans a wide range in energy, from $Q=4$~GeV to $Q=150$~GeV, and in $x$, from $x\sim0.5\cdot10^{-4}$ to $x\sim 1$.  We recall that a single DY data point is simultaneously sensitive to a larger and a smaller value of $x$. This is because the cross section is given by a pair of TMDPDFs, eq.~(\ref{def:xSec}), computed in $x_1$ and $x_2$ such that $x_1x_2 \simeq Q^2/s$, see eq.~(\ref{def:x12}).

In our fit we have compared absolute values of cross-section, whenever they are available. The only data set that require normalization factors are all CMS data, ATLAS at 7 TeV, and DO electron-pair measurements. For these sets we have normalized the integral of the theory prediction to corresponding integral over the data (see explicit expression in ref.\cite{Scimemi:2017etj}). To our best knowledge, it is the first fit of TMD factorization to absolute values of cross-section in the modern time, compare e.g to the latest and most advanced fits in \cite{DAlesio:2014mrz,Bacchetta:2017gcc,Scimemi:2017etj}.

\begin{figure}[t]
\begin{center}
\includegraphics[width=0.61\textwidth]{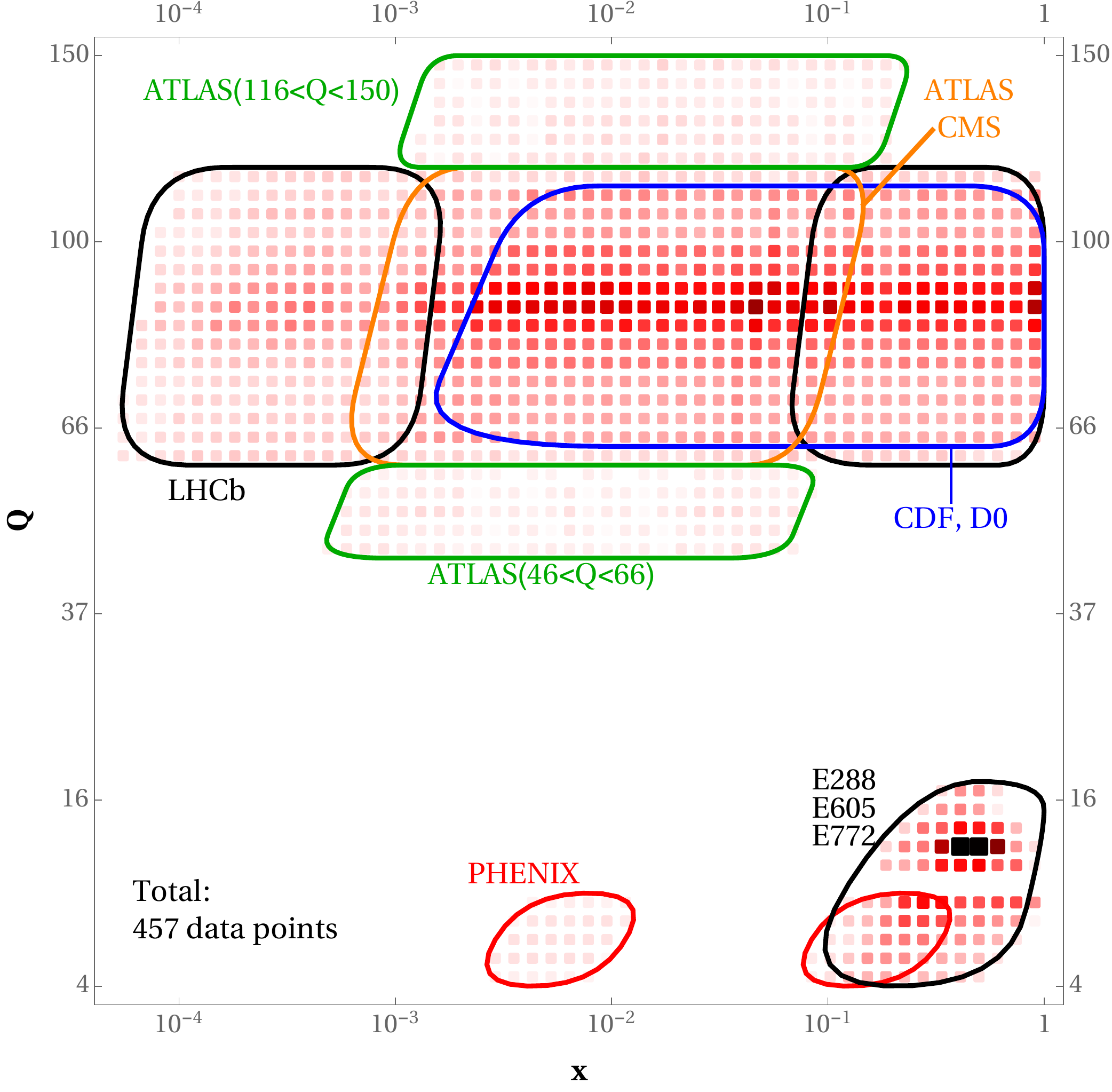}
\caption{\label{fig:dataPoints} Density distribution of data points in the plain $(Q,x)$ for  each experiment analyzed in the fit.}
\end{center}
\end{figure}
The kinematic region in $x$ and $Q$ covered by the data set considered for our fit is shown in fig.~\ref{fig:dataPoints}. The boxes enclose the sub-regions covered by the single data sets.  Looking at fig.~\ref{fig:dataPoints}, it is possible to distinguish two main clusters of data: the ``low-energy experiments'', \textit{i.e.} E288, E605, E772\footnote{Notice that the experiments E605 and E772 have been included in a fit of TMPDFs for the first time in this work.} and PHENIX, that place themselves at invariant-mass energies between 4 and 18~GeV, and the ``high-energy experiments'', \textit{i.e.} all those from Tevatron and LHC, that are instead distributed around the $Z$-peak region. From this plot we observe that, while the high-energy experiments span a wide range in $x$, the coverage in $x$ of the low-energy ones is more limited. This is a consequence of the fact all the low-energy experiments but PHENIX are fixed-target experiments. On the other hand, the number of data points belonging to the low-energy and high-energy experiments is of the same order ensuring a balanced distribution of data in $Q$.

\section{Statistical analysis}
\label{sec:stat}

In this section we discuss the treatment of the experimental information within our fit. The final purpose is to provide a suitable definition of the $\chi^2$ that allows for a correct exploitation of experimental uncertainties. A proper treatment of \textit{uncorrelated} and \textit{correlated} uncertainties is fundamental to obtain a faithful extraction of the TMDPDFs.

Let us consider an ensemble of $n$ measurements having the following
structure
\begin{equation}
  m_i\pm \sigma_{i,\rm stat} \pm \sigma_{i,\rm unc} \pm \sigma_{i,\rm
    corr}^{(1)}\pm\dots \pm \sigma_{i,\rm
    corr}^{(k)}\,,
\end{equation}
where $m_i$, with $i=1,\dots, n$, is the central value of the $i$-th measurement, $\sigma_{i,\rm stat}$ its (uncorrelated) statistical uncertainty, $\sigma_{i,\rm unc}$ its uncorrelated systematic uncertainty\footnote{There could be more than one uncorrelated   systematic uncertainty. In this case, $\sigma_{i,\rm unc}$ is just the square root of the sum in quadrature of all the uncorrelated   systematic uncertainties.}, and $\sigma_{i,\rm corr}^{(l)}$, with $l=1,\dots,k$, its correlated systematic uncertainties. Uncorrelated uncertainties give an estimate of the degree of knowledge of a particular data point irrespective of the other measurements of the data set. A typical example of uncorrelated uncertainty is the statistical one but also other systematic sources are possible. Correlated uncertainties, instead, provide an estimate of the correlation between the statistical fluctuations of two separate data points of the same data set. Typically, correlated uncertainties are of systematic origin, \textit{e.g.} they are connected with the apparatus used to perform the measurements.

With this information at hand, one can construct the experimental \textit{covariance matrix} $V_{ij}$ as follows (see for example ref.~\cite{Ball:2008by,Ball:2012wy}):
\begin{equation}\label{eq:covmat}
  V_{ij}=\left(\sigma_{i,\rm stat}^2 +\sigma_{i,\rm unc}^2\right)\delta_{ij} + \sum_{l=1}^{k}\sigma_{i,\rm
    corr}^{(l)}\sigma_{j,\rm
    corr}^{(l)}\,.
\end{equation}
Given a set of predictions $t_i$ corresponding to the $n$ measurements of the ensemble, the $\chi^2$ takes the form
\begin{equation}\label{eq:chi2cov}
  \chi^2=
  \sum_{i,j=1}^{n}\left(m_i-t_i\right)V_{ij}^{-1}\left(m_j-t_j\right) =
  \mathbf{y}^{T} \cdot \mathbf{V}^{-1} \cdot \mathbf{y}\,,
\end{equation}
where in the second equality we have used the matrix notation and defined the residuals $y_i = m_i-t_i$. The $\chi^2$ in eq.~(\ref{eq:chi2cov}) takes into account the possible different nature of the experimental uncertainties leading to a faithful estimate of the agreement between data and theoretical predictions. An efficient way to compute the $\chi^2$ in eq.~(\ref{eq:chi2cov}) is discussed in Appendix~\ref{app:cholesky}.

As we will show below, the presence of sizable correlated uncertainties may give rise to significant shifts such that a \textit{visual} comparison between central experimental values and theoretical predictions is misleading. Specifically, an apparent visual disagreement may still be compatible with an acceptable value of the $\chi^2$. However, it is possible to quantify the effect of the correlated uncertainties on the single data points by computing the so-called systematic shifts $d_i$. In this approach the $\chi^2$-value (\ref{eq:chi2cov}) is presented by a sum of two terms \cite{Ball:2012wy}
\begin{eqnarray}\label{eq1:chi2nuisshift}
\chi^2=\chi_D^2+\chi^2_\lambda,
\end{eqnarray}
where $\chi^2_D$ is the uncorrelated contribution and $\chi^2_\lambda$ is a penalty term. Loosely speaking, $\chi_D^2$($\chi^2_\lambda$) demonstrates the agreement in the shape(normalization) between theory and measurement. Applying these shifts to the theoretical predictions\footnote{They could be equally well applied to the experimental central values.} should produce a more trustful visual comparison. The explicit computation of the systematic shifts is presented in Appendix~\ref{app:sysshifts}.

\section{Results}
\label{sec:results}

In this section we present the results of our analysis. We  start commenting the quality of the fit and comparing the input data set to the theoretical predictions. Then we turn to consider the outcome for TMDPDFs and the numerical values of the parameters extracted from the fit. 
We detail our study on
error propagation from experimental data that  is handled by a Monte Carlo sampling, known also as the replica method. To this end, we have generated 100 pseudodata replicas according the rules described in ref.~\cite{Ball:2008by}, and we performed the $\chi^2$-minimization for each pseudodata set. The central values are the mean of the obtained 100 fits.

\subsection{Agreement between theory and experiment}
\label{sec:agree}

\begin{table}[t]
\small
\begin{center}
\begin{tabular}{|l|| c|c|c|c||c|}
\hline
Data set & $N_{\rm pt}$ & $\chi^2_D/N_{\rm pt}$ & $\chi^2_\lambda/N_{\rm pt}$ & $\chi^2/N_{\rm pt}$& $\langle d/\sigma\rangle$  
\\
\hline \hline
E288 (200)				&	43	&	0.79	&	0.06	&	0.86 &	$41.15\%$	
\\
\hline
E288 (300)				&	53	&	0.89	&	0.04	&	0.93 &	$35.72\%$	
\\
\hline
E288 (400) 				&	76	&	0.78	&	0.01	&	0.80 &	$26.52\%$	
\\
\hline
E605 					&	53	&	0.49	&	0.05	&	0.54 &	$24.74\%$	
\\
\hline
E772					&	35	&	1.65	&	0.05	&	1.70 &	$13.24\%$	
\\
\hline
PHENIX 					&	3	&	0.28	&	0.02	&	0.30	&	$4.08\%$	
\\
\hline\hline
\textbf{Low energy data}	&	\textbf{263}	&	\textbf{0.86}	&	\textbf{0.04}	&	\textbf{0.90} &			
\\
\hline\hline
CDF (run1)					&	33	&	0.54	&	0.14	&	0.68 &	$8.42\%$	
\\
\hline
CDF (run2)					&	39	&	1.37	&	0.01	&	1.37 &	$2.90\%$	
\\
\hline
D0 (run1)					&	16	&	0.76	&	0.00	&	0.76	&	$0.12\%$	
\\
\hline
D0 (run2)					&	8	&	1.51	&	0.00	&	1.51 &	$0.00\%$	
\\
\hline
D0 (run2)$_\mu$				&	3	&	0.33	&	0.36	&	0.68 &	$0.33\%$	
\\
\hline \hline
Tevatron				    &	99	&	0.97	&	0.06	&	1.03 &			
\\
\hline\hline
ATLAS (7 TeV) $|y|<1$	&	5	&	2.16	&	0.00	&	2.17 &	$-0.05\%$	
\\
\hline
ATLAS (7 TeV) $1<|y|<2$		&	5	&	5.13	&	0.00	&	5.14 &	$-0.07\%$
\\
\hline
ATLAS (7 TeV) $2<|y|<2.4$	&	5	&	1.08	&	0.00	&	1.08	&	$-0.02\%$	
\\
\hline \hline
ATLAS (8 TeV) $|y|<0.4$		&	5	&	1.86	&	0.33	&	2.19 &	$3.68\%$
\\
\hline
ATLAS (8 TeV) $0.4<|y|<0.8$	&	5	&	2.41	&	0.68	&	3.09 &	$3.66\%$	
\\
\hline
ATLAS (8 TeV) $0.8<|y|<1.2$	&	5	&	1.02	&	0.54	&	1.56 &	$3.77\%$	
\\
\hline
ATLAS (8 TeV) $1.2<|y|<1.6$	&	5	&	1.24	&	0.49	&	1.73 &	$4.29\%$	
\\
\hline
ATLAS (8 TeV) $1.6<|y|<2.0$	&	5	&	0.42	&	0.59	&	1.01 &	$4.93\%$	
\\
\hline
ATLAS (8 TeV) $2.0<|y|<2.4$	&	5	&	1.55	&	1.21	&	2.76 &	$5.56\%$	
\\
\hline
ATLAS (8 TeV) 46 - 66 GeV	&	3	&	0.43	&	0.07	&	0.49 &	$1.45\%$
\\
\hline
ATLAS (8 TeV) 116 - 150 GeV	&	7	&	0.74	&	0.13	&	0.87 &	$1.96\%$
\\
\hline\hline
ATLAS total					&	55	&	1.65	&	0.37	&	2.02 &			
\\
\hline\hline
CMS (7 TeV)					&	8	&	1.26	&	0.00	&	1.26 &	$0.00\%$	
\\
\hline
CMS (8 TeV)					&	8	&	0.85	& 	0.00	&	0.85 &	$0.00\%$	
\\
\hline\hline
CMS total					&	16	&	1.06	&	0.00	&	1.06 &			
\\
\hline\hline
LHCb (7 TeV)				&	8	&	2.05	&	0.90	&	2.95 &	$5.69\%$	
\\
\hline
LHCb (8 TeV)				&	7	&	3.85	&	1.69	&	5.54 &	$5.65\%$	
\\
\hline
LHCb (13 TeV)				&	9	&	0.60	&	0.29	&	0.89 &	$6.34\%$	
\\
\hline\hline
LHCb total				&	24	&	2.03	&	0.90	&	2.93 &			
\\
\hline\hline
\textbf{High energy data}     &	\textbf{194}	&	\textbf{1.30}	&	\textbf{0.25}	&	\textbf{1.55} &			
\\
\hline\hline
  \textbf{Global}             	        &	\textbf{457}	&	\textbf{1.05}	&	\textbf{0.12}	&	\textbf{1.17} &			
\\\hline
\end{tabular}
\caption{\label{tab:resultCHI} Distribution of values of $\chi^2$ over the data set. Decomposition of $\chi^2$ to uncorrelated part $\chi^2_D$ and shift part $\chi^2_\lambda$ is made with nuisance parameter. The average shift is (resulted from the nuisance parameters) is shown relative to the value of cross section. }
\end{center}
\end{table}

In tab.~\ref{tab:resultCHI} we report the values of the $\chi^2$ (for central values), normalized to the number of data points $N_{\rm pt}$, for the individual experiments, for some relevant subsets of experiments, and for the global data set included in this analysis. Specifically,  tab.~\ref{tab:resultCHI}  displays, along the number of data points $N_{\rm pt}$, the uncorrelated contribution to the $\chi^2$ ($\chi_D^2$), the penalty term ($\chi_\lambda^2$), and the sum of the two, \textit{i.e.} the total $\chi^2$  referring to eq.~(\ref{eq1:chi2nuisshift}) (see also eq.~(\ref{eq:chi2nuisshift})). The last column, instead, reports the average (over the data set) systematic shift $d_i$ (as defined in eq.~(\ref{eq:sysshiftdef})), over the cross-section value in percentage.

The first observation is that the value of the global $\chi^2$ is particularly good ($\chi^2/N_{\rm pt} = 1.18$). This means that the fit has achieved a satisfactory description of the entire data set. We also observe that the description of the low-energy subset is substantially better ($\chi^2/N_{\rm pt} = 0.93$) than the high-energy one ($\chi^2/N_{\rm pt} = 1.52$). This is not surprising because the high-energy experiments from Tevatron and LHC are much more accurate than the low-energy ones. In addition, amongst the high-energy experiments, LHCb has the largest $\chi^2$, while ATLAS, CMS, and the Tevatron experiments are fairly described. Dropping the best (PHENIX) and the worst (LHCb 8TeV) set (in total 10 points), we get $\chi^2/N_{\text{pt}}=1.12$.

\begin{figure}[t]
\begin{center}
\includegraphics[width=0.95\textwidth]{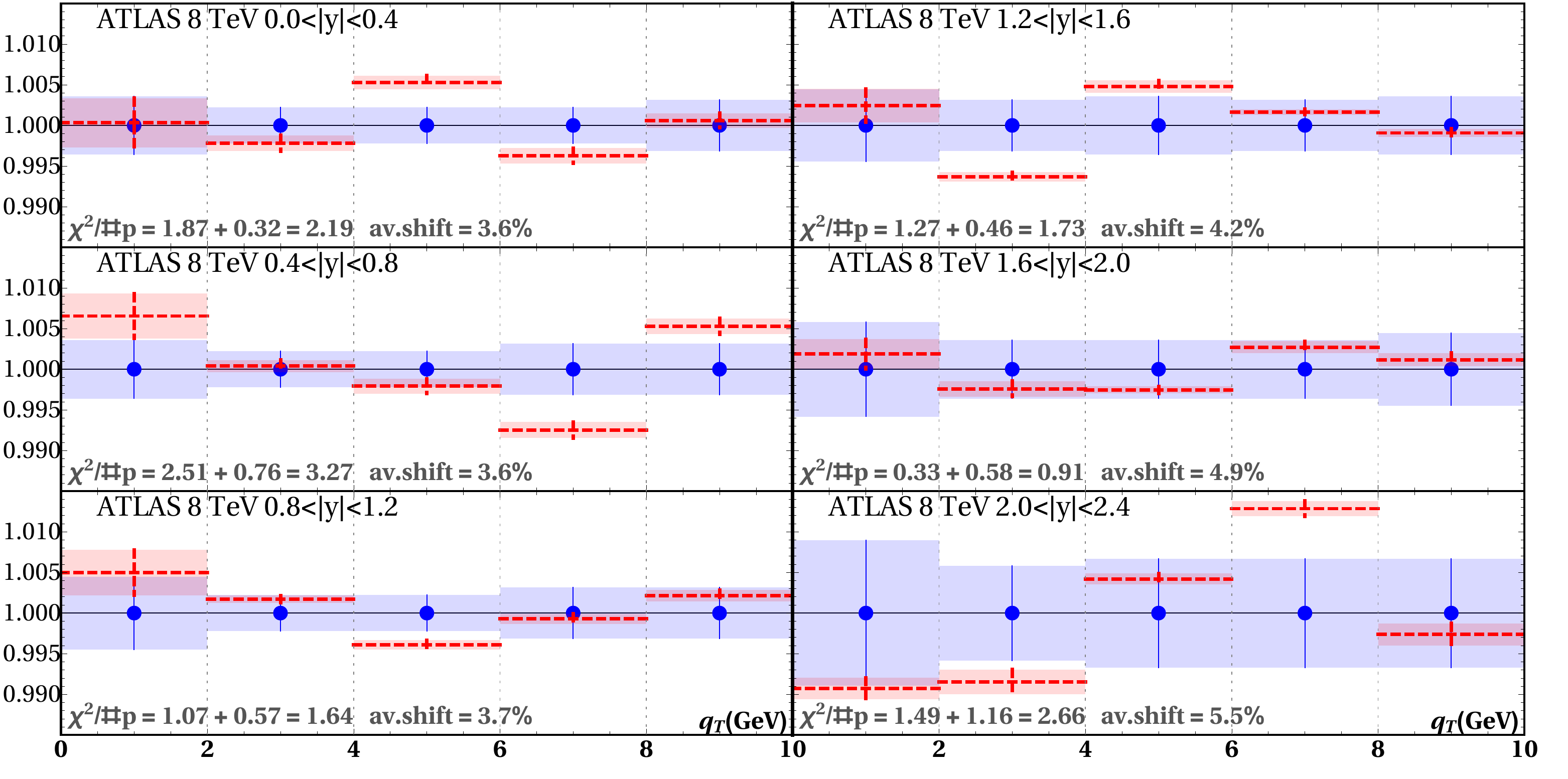}
\caption{\label{fig:atlas}Ratio of theoretical and experimental points as a function of the binned di-lepton transverse momentum for the measured at ATLAS in the range $66<Q<116$ GeV (dashed red lines). The experimental points (blue dots) are surrounded by a box describing their error. The representation takes into account the shifts as described in the text. }
\end{center}
\end{figure}
\begin{figure}[t]
\begin{center}
\includegraphics[width=0.45\textwidth]{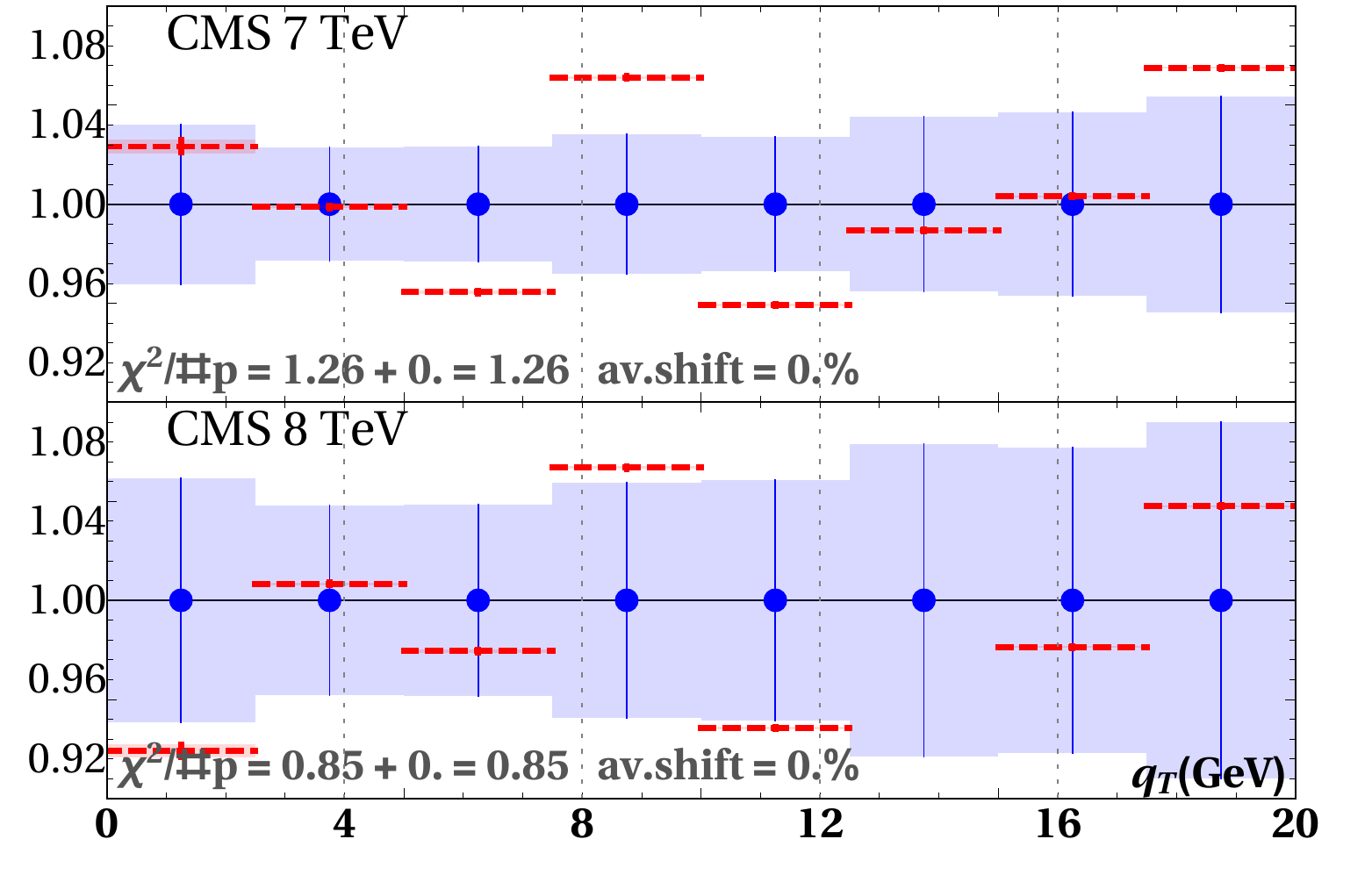}
~~
\includegraphics[width=0.45\textwidth]{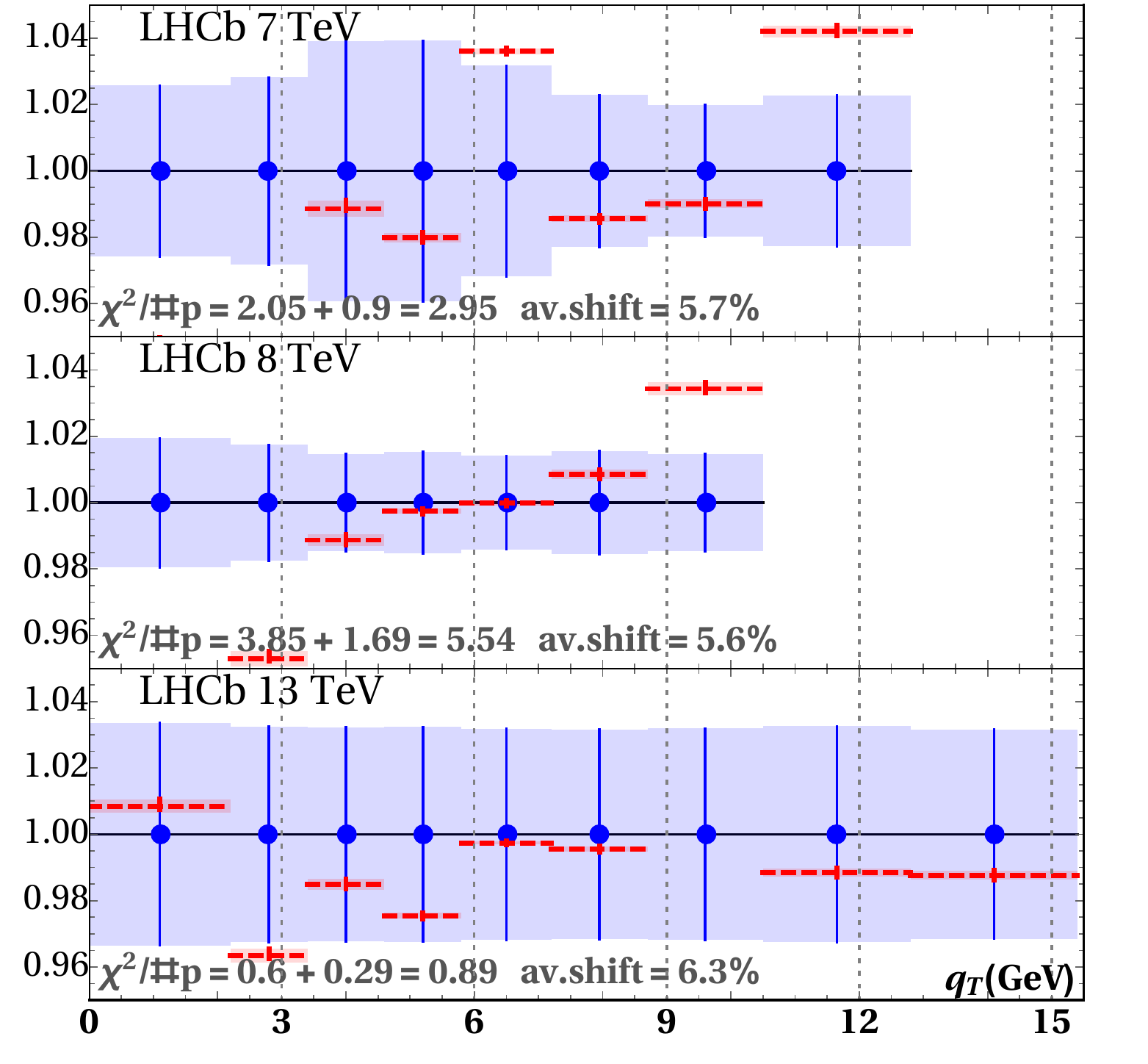}
\caption{\label{fig:CMS+LHCb}Ratio of theoretical and experimental points as a function of the binned di-lepton transverse momentum for the measured at CMS and LHCb experiments (dashed red lines). The experimental points (blue dots) are surrounded by a box describing their error. The representation takes into account the shifts as described in the text.}
\end{center}
\end{figure}

\begin{figure}[t]
\begin{center}
\includegraphics[width=0.9\textwidth]{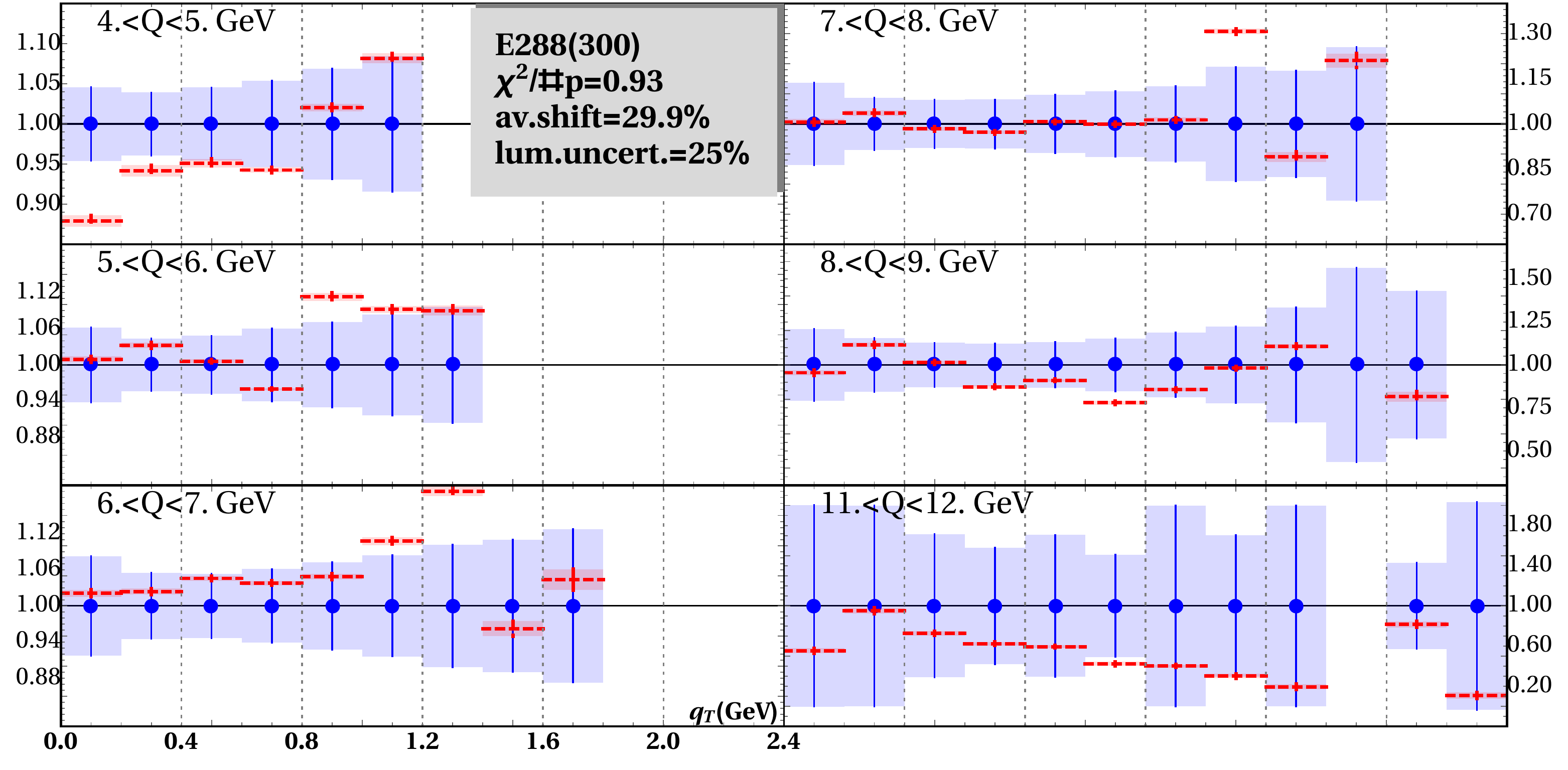}\\
\caption{\label{fig:e300} Ratio of theoretical and experimental points as a function of the binned di-lepton transverse momentum at  E288 (300) (dashed red lines). The experimental points (blue dots) are surrounded by a box describing their error. The representation takes into account the shifts as described in the text.}
\end{center}
\end{figure}

In order to achieve a visual assessment of the agreement between data and theory, in fig.~\ref{fig:atlas}, \ref{fig:CMS+LHCb}, \ref{fig:e300} we display the ratio between theoretical predictions (red dashed lines) and experimental data points along with their uncorrelated uncertainty (blue bands) for some representative data sets included in the fit. In particular,  we show plots for the LHC and one of the E288 data sets. 
An example of cross-section values without systematic shifts is given in appendix~\ref{app:sysshifts} in fig.~\ref{fig:atlas+}. The theoretical predictions have been corrected including the systematic shifts computed as described in Appendix~\ref{app:sysshifts} (see eq.~(\ref{eq:shiftedpreds})).

From fig.~\ref{fig:atlas}-\ref{fig:CMS+LHCb}, we see that, despite the small experimental uncorrelated uncertainties at the percent level or below, our fit is able to describe the LHC data sets fairly well. However, the 8~TeV data set of LHCb presents a pronounced shape discrepancy that causes the large value of the $\chi^2$ reported in tab.~\ref{tab:resultCHI}. A similar tension between data and theory seems to be present also in the most forward rapidity bin ($2<|y|<2.4$) of the ATLAS data set at 8~TeV. We ascribe the origin of the discrepancy to the insufficient shape of collinear PDFs at very large $x$ ($x\simeq 0.7$). In this region, collinear PDFs are poorly known. The fact that TMDPDF is sensitive to the shape of collinear PDF could be used to constrain the behavior of PDF. Such a study is certainly interesting but goes beyond the scope of this paper. Note, that the LHCb set could also be affected by the poor knowledge of PDFs at small-$x$, since for this set $x$ reaches values down to $\sim 10^{-4}$.

In fig.~\ref{fig:e300}, the data-theory comparison for one of the E288 data sets shows that the uncorrelated experimental uncertainties range between 5\% and a few tens of percent.  Such large uncertainties make the agreement with the theoretical predictions easier to achieve, giving rise to small $\chi^2$'s. Similar comments apply to all low energy experiments. We note the systematic underestimation for the cross-section for experiments E288, E605 and E772, which is of the order of $25\%$ on average. Nonetheless, such a large difference between data and the theory does not produce large $\chi^2$-values, due to large systematic uncertainties for this data. The reported correlated systematic error for E288(E605, E772) experiments is 25\%(15\%, 10\%) \cite{Ito:1980ev,Moreno:1990sf,McGaughey:1994dx}. This systematic discrepancy has been recently discussed in \cite{Bacchetta:2019tcu}, where it was connected to the fixed-target nature of these experiments.

\subsection{Extracted values of TMDPDF and rapidity anomalous dimension}
\label{sec:values}

We now turn to the  values of the TMDPDFs and rapidity anomalous dimension as extracted from the fit. Our results for the non-perturbative parameters are presented in tab.~\ref{tab:result}. The central values and the uncertainty band correspond to the mean and standard deviation of parameter distributions obtained by $\chi^2$-minimization of 300 pseudodata replicas. One should take into account that the uncertainties presented here take into account the correlation among parameters.

\begin{table}[b] \small 
\begin{center} 
\begin{tabular}{|c|c||c|c|c|c|c|}
\hline
 $B_{\text{NP}}$ & $c_0$ & $\lambda_1$ & $\lambda_2$ & $\lambda_3$ & $\lambda_4$ & $\lambda_5$ 
\\\hline
\multicolumn{7}{|c|}{\textbf{Full data set}}
\\\hline
$3.31\pm 0.28$ & $0.024 \pm 0.006$ & $0.258 \pm 0.022$ & $8.18 \pm 1.00$ & $-4.76 \pm 1.38$ & $300. \pm 89.$ & $2.44 \pm 0.12$
\\\hline
$2.5$(fixed) & $0.037 \pm 0.007$ & $0.248 \pm 0.025$ & $8.15 \pm 1.40$ & $-4.96 \pm 1.60$ & $275. \pm 53.$ & $2.52 \pm 0.13$
\\\hline
\multicolumn{7}{|c|}{\textbf{Excluding LHC-data}}
\\\hline
$1.21\pm 0.50$ & $0.057 \pm 0.038$ & $0.21 \pm 0.17$ & $12.1 \pm 4.4$ & $-3.51 \pm 5.40$ & $316. \pm 196.$ & $2.11 \pm 0.28$
\\\hline
$2.5$(fixed) & $0.014 \pm 0.012$ & $0.14 \pm 0.08$ & $11.2 \pm 3.8$ & $-2.48 \pm 3.96$ & $413. \pm 277.$ & $2.07 \pm 0.21$
\\\hline
\end{tabular}
\caption{\label{tab:result} 
Values of parameters extracted in the fit in the model (\ref{model:exp},~\ref{model:our_fNP}). The error corresponds to a standard uncorrelated deviation calculated over 300(100) replicas for full(reduced) data set.}
\end{center}
\end{table}

Analyzing the result of the fit one should  keep in mind that high-energy and low-energy experiments unequally contribute to the $\chi^2$-value.  Because the data from LHC have  tiny errors (especially the data measured at ATLAS), they contribute decisively  to the value of $\chi^2$. For this reason, the minimum of $\chi^2$ is shifted  towards the local minimum of the LHC data set, especially for  smaller $x$  values (say $x\lesssim 0.05$). To determine the effect of the LHC data set we have additionally performed a fit without the LHC data (100 pseudodata replicas) and we show the results in the second part of tab.~\ref{tab:result}. One can see that the values obtained in both fits nicely agree with each other, apart from  for $B_{\text{NP}}$ (and we discuss this fact later in the text). It is clear that inclusion of the LHC data  affects very strongly the uncertainty in the parameter determination.

\begin{figure}[t]
\begin{center}
\includegraphics[width=0.35\textwidth]{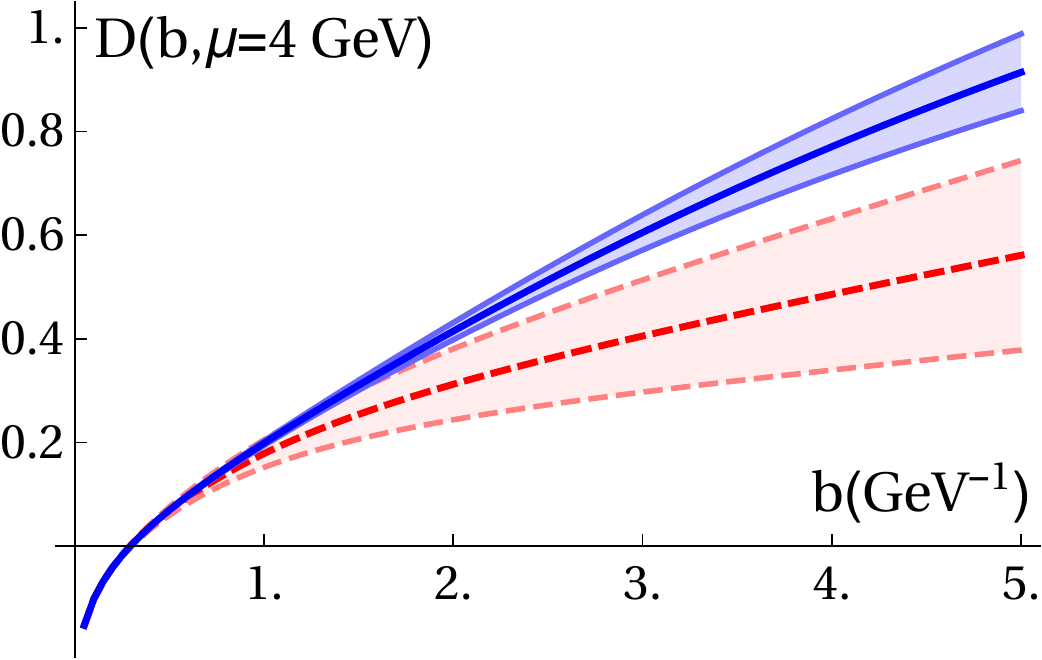}~~
\includegraphics[width=0.35\textwidth]{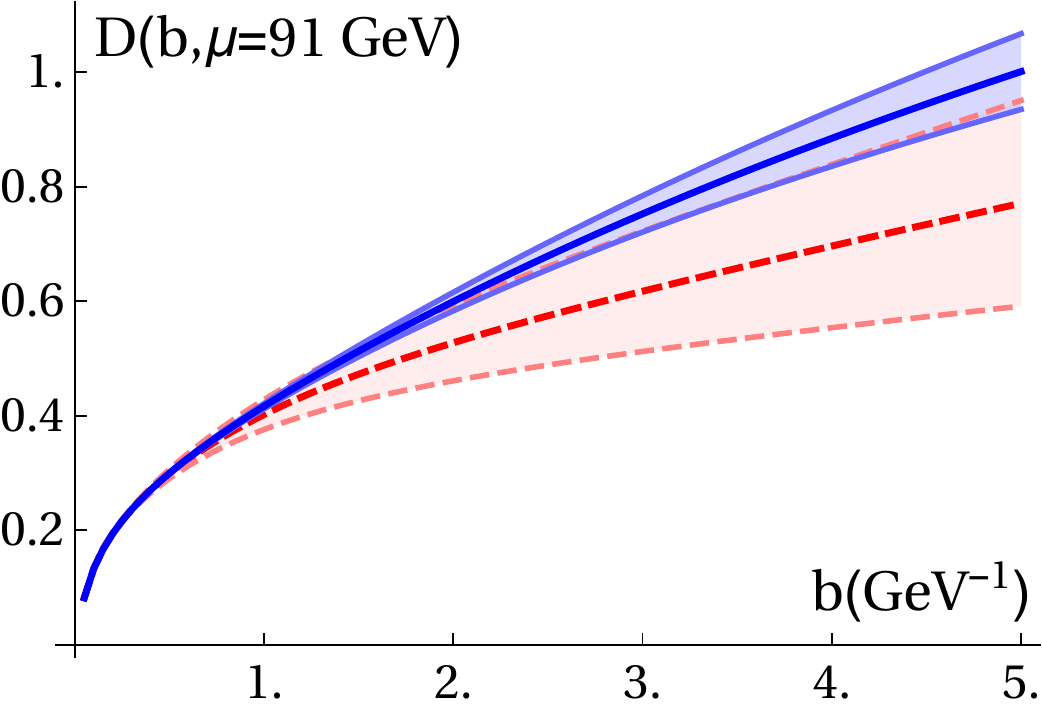}
\caption{\label{fig:DbT}The $\mathcal{D}$ anomalous dimension in $b$ space   for two values of $\mu$. The bands correspond  respectively to the  case in which one  includes all experiments  (blue) and to the case in which  LHC data are excluded (red-dashed).}
\end{center}
\end{figure}

The plot of the extracted rapidity anomalous dimension (together with $1\sigma$ band) is shown in fig.~\ref{fig:DbT} at $\mu=4$ GeV and $\mu=91$ GeV. One can see that the fitted value of $B_{\text{NP}}$ is pretty large. This reflects the fact that  high-energy experiments (which dominate our $\chi^2$) prefer the entirely perturbative rapidity anomalous dimension. This was already pointed out in previous works~\cite{Echevarria:2012pw,DAlesio:2014mrz,Scimemi:2017etj}. The value of the parameter $c_0$ extracted from the fit is compatible with the renormalon approximation discussed in ref.~\cite{Scimemi:2016ffw}. In the absence of LHC measurements the fitted value of $B_{\text{NP}}=1.2$, which is very close to values obtained in previous LHC-less data fits (compare to $b_{\text{max}}\sim 1.1$ in ref.~\cite{Landry:2002ix,Bacchetta:2017gcc}). 

We have observed that the values of global $\chi^2$ (for the full data set) are practically the same for the values of $B_{\text{NP}}$ in a wide region. Fixing $B_{\text{NP}}=\{1.,2.,3.,4.\}$ GeV$^{-1}$ we have obtained the minimal values of $\chi^2/N_{\text{pt}}=\{1.27,1.18,1.17,1.18\}$. At larger $B_{\text{NP}}$, the fit becomes unstable due to influence of the Landau pole (the actual position of the singularity in the resummed expression depends on the realization of the strong coupling values at very-low energies, and typically located at $b=5.-8.$ GeV$^{-1}$.). We admit that the distribution of the $\chi^2$ between experiments is different. In particular, for the very large $B_{\text{NP}}$ small-value of $\chi^2$ is achieved by better agreement with LHCb experiment, whereas the agreement with the majority of the data is worsen. Considering this picture, we conclude that the obtained error-band on $B_{\text{NP}}$, presented in table \ref{tab:result}, does not reflect the realistic state. It is probably due to strong correlation between $B_{\text{NP}}$ and other parameters, and due to the theory-data tension for some particular data subsets. To support the extraction presented here, and to show that it is not strongly affected by this freedom, we have also performed the fit of the data at fixed $B_{\text{NP}}=2.5 $ GeV$^{-1}$.  The results are presented in table \ref{tab:result}. Clearly, all parameter of $f_{\text{NP}}$ are in agreement within uncertainty band, while the value of $c_0$ (which is anti-correlated to $B_{\text{NP}}$ is tends to compensate its change.

\begin{figure}[t]
\begin{center}
\includegraphics[width=0.32\textwidth]{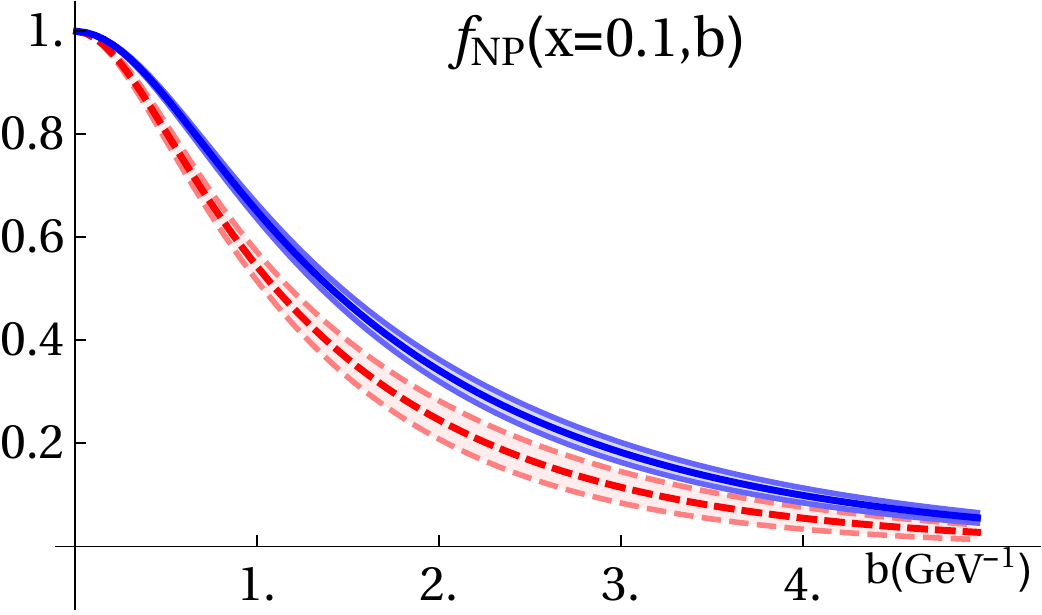}
\includegraphics[width=0.32\textwidth]{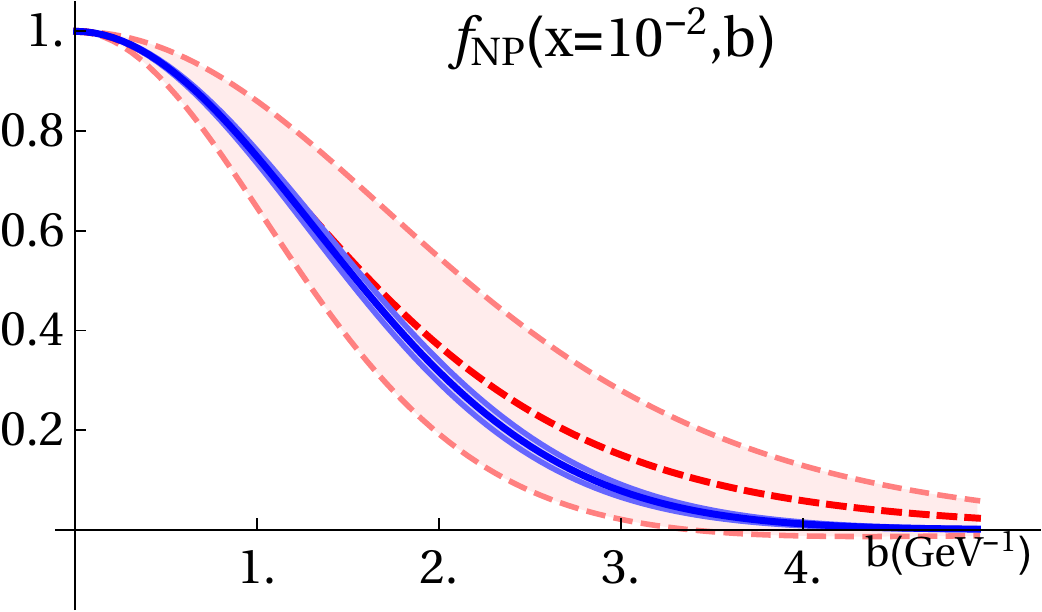}
\includegraphics[width=0.32\textwidth]{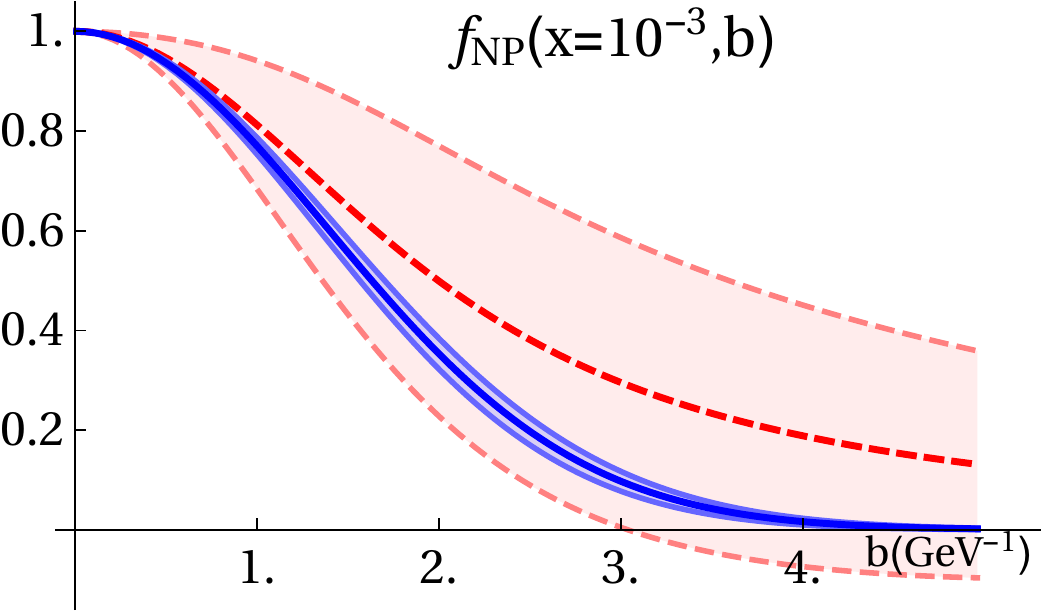}
\caption{\label{fig:FNPbT}The intrinsic non-perturbative part of the TMDPDF as  in eq.~(\ref{model:our_fNP}). The bands correspond  respectively to the  case in which one  includes all experiments  (blue) and to the case in which  LHC data are excluded (red-dashed).}
\end{center}
\end{figure}

In fig.~\ref{fig:FNPbT} we show the intrinsic non-perturbative part of TMDPDF, $f_{NP}$, as a function of $\vec b$ at different values of $x$. 
We present  $f_{NP}$ extracted respectively from the full  (blue band) and from LHC-less (red band) data sets.
 Notably and for all the values of $x$, the inclusion of LHC data reduces the error-band. The reduction is not so significative  at $x\sim 0.1$, but it is an order of magnitude at $x\sim 10^{-3}$. 
 One should also take into account that this picture is somewhat model-biased. The high-energy experiments (and thus LHC data) are sensitive to small-$b$ values (say $b\lesssim 2$ GeV$^{-1}$) and they are practically insensitive to  large-$b$ values.
  On the contrary  for the low-energy experiments one finds that the values of $b\sim 5$-$6 $ GeV$^{-1}$ give a sizable contribution to the cross-section. Given the small number of parameters in our model, one cannot entirely decorrelate large and small $b$ behavior, and thus the error-band at large-$b$ is particularly underestimated.

\begin{figure}[t]
\begin{center}
\includegraphics[width=0.35\textwidth]{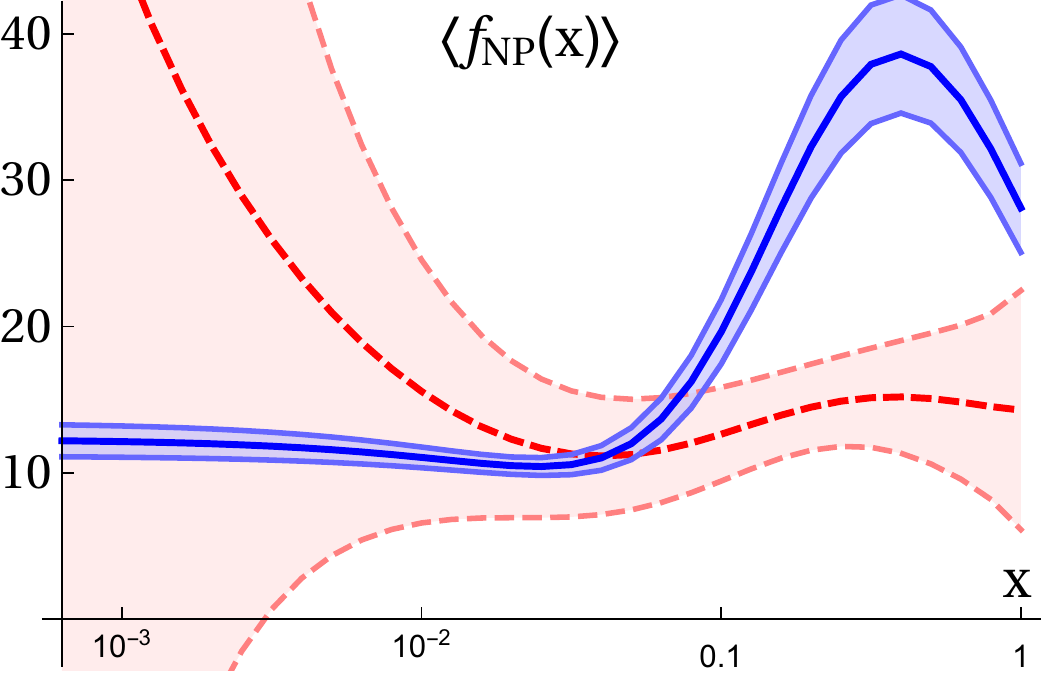}
\includegraphics[width=0.35\textwidth]{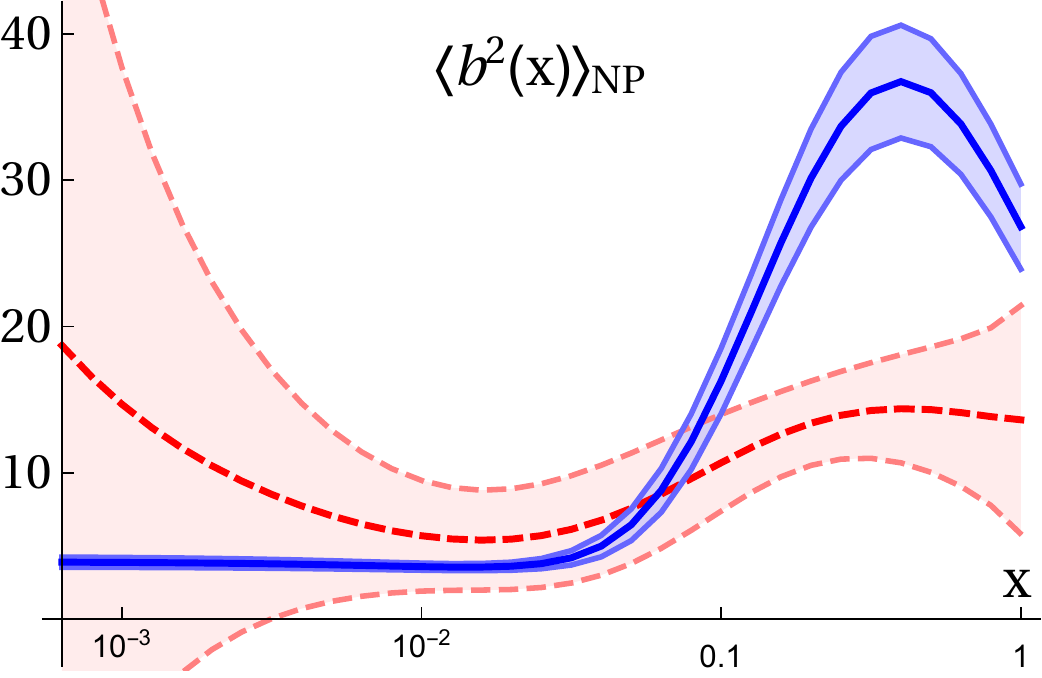}
\caption{\label{fig:avb2} The moments of $f_{NP}$ defined in (\ref{moments}) as a function of $x$. The blue (red-dashed) bands correspond to extraction made with (without) LHC data.}
\end{center}
\end{figure}

An important feature of our extraction is the essential dependence of  $f_{NP}$ on $x$. Indeed, in the overwhelming part of previous studies (see e.g.~\cite{Landry:2002ix,DAlesio:2014mrz,Scimemi:2017etj}) the $x$ dependence of  $f_{NP}$ was absent (an exception is the $x$-dependent $f_{NP}$ in ref.~\cite{Bacchetta:2017gcc}). In our case, the $x$-dependence is strong and it has been uncovered due to presence of high-precision high-energy experiments. We have checked, that we are not able to fit LHC data with $x$-independent $f_{NP}$, whereas the rest of data could  equally-well be described by a simpler $x$-independent $f_{NP}$. We have found that the present data set prefers a wide exponential-like $f_{NP}$ at larger $x$ ($x\sim 0.1-0.5$) and narrower Gaussian-like $f_{NP}$ at smaller $x$. In order to quantify this behavior we consider $b$-moments of  $f_{NP}$ defined as
\begin{eqnarray}\label{moments}
\langle f_{NP}(x)\rangle = \int d^2 \vec b \,f_{NP}(x,\vec b),
\qquad\qquad
\langle b_{NP}^2(x)\rangle = \frac{\int d^2 \vec b\, \vec b^2\, f_{NP}(x,\vec b)}{\langle f_{NP}(x)\rangle}.
\end{eqnarray}
The values of $\langle f_{NP}(x)\rangle$ and $\langle b_{NP}^2(x)\rangle$ are shown in fig.~\ref{fig:avb2}. Unfortunately, these functions have no direct physical meaning, but they  
show clearly that  at $x\gtrsim 0.05$ the non-perturbative behavior of the unpolarized TMDPDF changes to become wider and exponential-like. 
In  $k_T$-space it would correspond to a narrower $k_T$ distribution for larger $x$. Such behavior has been already observed in ref.~\cite{Bacchetta:2017gcc}. 
Still observing fig.~\ref{fig:avb2}, it is clear that the data without LHC points have no restricting power for $x\lesssim 10^{-2}$.

\begin{figure}[t]
\begin{center}
\includegraphics[width=0.45\textwidth]{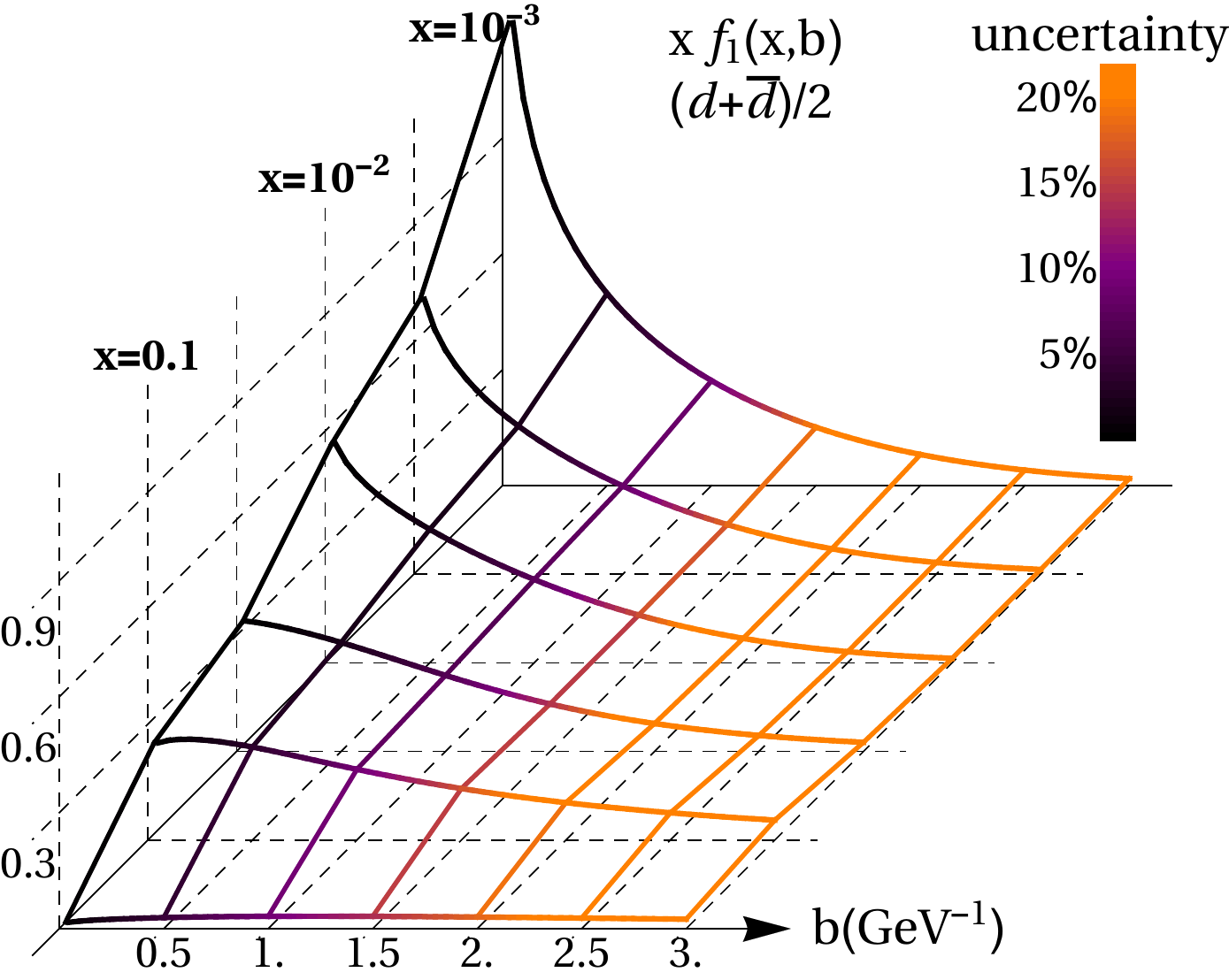}~~
\includegraphics[width=0.45\textwidth]{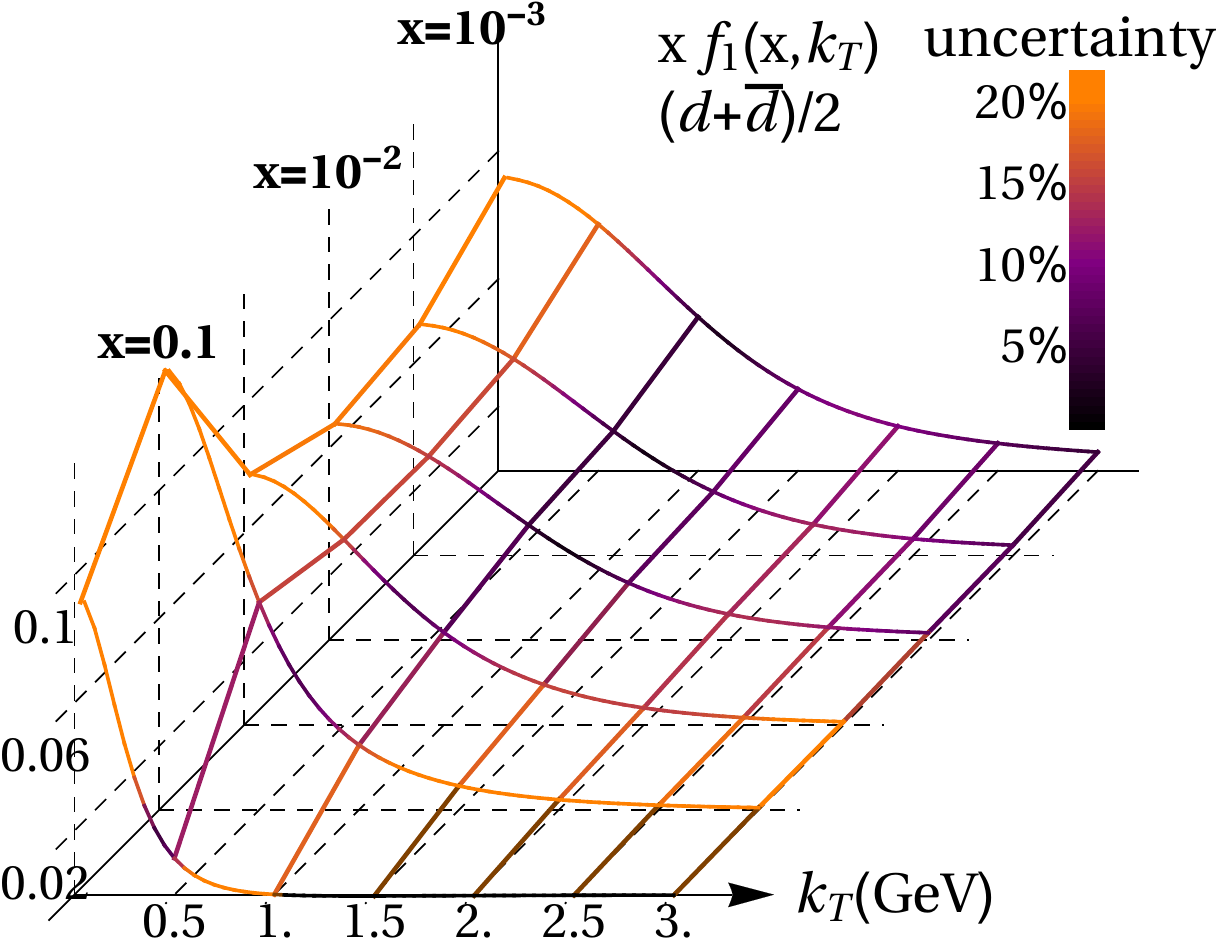}
\caption{\label{fig:fbT}The down quark TMD PDF in $b$-space(left) and $k_T$-space(right) presented at different values of $x$. The color shows the size of the uncertainty relative the value of distribution.}
\end{center}
\end{figure}

Finally, in fig.~\ref{fig:fbT} we present the three-dimensional illustration for the unpolarized TMDPDF $f_1$ in position and momentum spaces. The TMDPDF in  momentum space is defined as
\begin{eqnarray}
f_1(x,\vec k_T)=\int \frac{d^2\vec b}{(2\pi)^2}f_1(x,\vec b)e^{-i(\vec b\cdot \vec k_T)}.
\end{eqnarray}
The 1$\sigma$-uncertainty level is presented by color since the absolute value of the band is visually unresolved. 
For demonstration purposes we present the combination of the $d$- and $\bar d$-flavor distributions. Note, that generally, $f_{NP}$ is flavor dependent, although we omit its flavor dependence in the present work. Nonetheless, the extracted TMDPDFs have a flavor dependence and  it is driven solely  by the collinear PDF. The results of the extraction, together with the code for the cross-section, are available as a part of the \texttt{artemide} package \cite{web}. The replicas of full data set and LHC-less data set are labeled as \texttt{BSV19.bFIT} and \texttt{BSV19.bFIT.noLHC} correspondingly. The extractions with the fixed $B_{\text{NP}}=2.5$ GeV$^{-1}$ are labeled by \texttt{BSV19.bFIX} and \texttt{BSV19.bFIX.noLHC}.

\section{Conclusions}

We have  extracted  the unpolarized transverse momentum dependent parton distribution function (TMDPDF) and rapidity anomalous dimension (also known as Collins-Soper kernel) from Drell-Yan data. The analysis has been performed in the $\zeta$-prescription with NNLO perturbative inputs. 
We have also provided an estimation of the errors on the extracted functions with the replica method. The values of TMDPDF and rapidity anomalous dimension, together with the code that evaluates the cross-section, are available at \cite{web}, as a part of the \texttt{artemide} package. We plan to release grids for TMDPDFs extracted in this work also through the TMDlib \cite{Hautmann:2014kza}.

Theoretical predictions are based on the newly developed concepts of $\zeta$-prescription and optimal TMD proposed in ref.~\cite{Scimemi:2018xaf}. This combination provides a clear separation between the non-perturbative effects in the evolution factor and the intrinsic transverse momentum dependence. Additionally, the $\zeta$-prescription permits the usage of different perturbative orders in the collinear matching and TMD evolution. For that reasons, the precise values of the rapidity anomalous dimension ($\pm 1\%(4\%,6\%)$ accuracy at $b=1(3,5)$ GeV$^{-1}$) are relevant for any observable that obeys TMD evolution.

In our analysis, we have included a large set of data points, which spans a wide range of energies ($4<Q<150$ GeV) and $x$ ($x>10^{-4}$), see fig.~\ref{fig:dataPoints}. The data set can be roughly split into the low-energy data, which includes experiments E288, E605, E772 and PHENIX at RHIC, and the high-energy data from Tevatron (CDF and D0) and LHC (ATLAS, CMS, LHCb) in similar proportion. 
To exclude the influence of power corrections to TMD factorization we consider only the low-$q_T$ part of the data set, as described in sec.~\ref{sec:data}. 
A good  portion of  data is included in the fit of TMD distributions for the first time, that is the data from E772, PHENIX, some parts of ATLAS and D0 data. 
For the first time, the data from LHC have been included without restrictions (the only previous attempt to include LHC data in a  TMDPDF fit is~\cite{Scimemi:2017etj}, where systematic uncertainties and normalization has been treated in a simplified manner). 
We have shown that the inclusion of LHC data greatly restricts the non-perturbative models at smaller $b$ ($b\lesssim 2$ GeV$^{-1}$) and smaller $x$ ($x\lesssim 0.05$), and therefore
they are  highly relevant for studies of the intrinsic structure of hadrons. 
A detailed comparison of fits with and without LHC data has been discussed in sec.~\ref{sec:results}.

The extracted TMDPDF shows a non-trivial $x$-dependence that is not dictated only by the collinear asymptotic limit of PDFs. 
In particular, we find that the unpolarized TMDPDF is bigger (in impact parameter space) at larger $x$, see fig.~\ref{fig:avb2}. 
This indirectly implies a smaller  value of the typical parton transverse momentum $k_T$ for larger $x$. 
A similar behavior has been also observed in~\cite{Bacchetta:2017gcc}. 
We also find a strong dependence on the PDF set. The PDFs play the role of a "model-independent" input at small values of $b$, and largely determines the $x$-dependence of TMDPDF. In particular, we have used the NNPDF3.1(nnlo) set~\cite{Ball:2017nwa}, since it provides the best agreement with data.
 We think that the reason for the better agreement with this PDF set is that it has been fitted to the modern LHC data. The fact that TMD observables are so sensitive to the collinear input can be used to put extra restrictions to PDFs. A detailed study of this possibility is  left for the future.

\acknowledgments 
A.V. thanks A. Prokudin and N. Sato for stimulating discussions.  V.B. acknowledges support from the European Research Council (ERC) under the European Union's Horizon 2020 research and innovation program (grant agreement No. 647981, 3DSPIN). I.S. is supported by the Spanish MECD grant FPA2016-75654-C2-2-P. 
 
\appendix

\section{Efficient computation of $\chi^2$}
\label{app:cholesky}

The evaluation of $\chi^2$ values (\ref{eq:chi2cov}) involves the inversion of voluminous covariance matrix. A convenient way to compute the $\chi^2$ relies on the Cholesky decomposition of the covariance matrix $\mathbf{V}$, which is presented in this appendix.

The Cholesky decomposition can be applied for any symmetric and positive definite matrix, such as the covariance matrix $\mathbf{V}$, defined in eq.~(\ref{eq:covmat}). 
The decomposition has the form
\begin{equation}\label{eq:choleskydec}
\mathbf{V} = \mathbf{L}\cdot\mathbf{L}^{T}\,,
\end{equation}
where $\mathbf{L}$ is a lower triangular matrix whose entries are related recursively to those of $\mathbf{V}$ as follows:
\begin{equation}\label{eq:cholalg}
\begin{array}{rcl}
  L_{kk} &=&\displaystyle \sqrt{V_{kk}-\sum_{j=1}^{k-1}L_{kj}^2}\,,\\
  \\
  L_{ik} &=&\displaystyle
             \frac{1}{L_{kk}}\left(V_{ik}-\sum_{j=1}^{k-1}L_{ij}L_{kj}\right)\,,\quad
             k < i\,,\\
\\
  L_{ik} &=&\displaystyle 0\,,\quad
             k > i\,.\\
\end{array}
\end{equation}
It is then easy to see that the $\chi^2$ can be written as
\begin{equation}
\chi^2 = \left|\mathbf{L}^{-1}\cdot \mathbf{y}\right|^2\,.
\end{equation}
Now, the vector $\mathbf{x} \equiv \mathbf{L}^{-1}\cdot \mathbf{y}$ is
the solution of the lower-diagonal linear system:
\begin{equation}
  \mathbf{L} \cdot \mathbf{x} = \mathbf{y}\,,
\end{equation}
that can be efficiently solved by forward substitution, so that:
\begin{equation}
  \chi^2 = \left|\mathbf{x}\right|^2\,.
\end{equation}
Following this procedure, one does not need to compute explicitly the
inverse of the covariance matrix $\mathbf{V}$, simplifying
significantly the computation of the $\chi^2$.

\section{Determining the systematic shifts}
\label{app:sysshifts}

\begin{figure}[t]
\begin{center}
\includegraphics[width=0.95\textwidth]{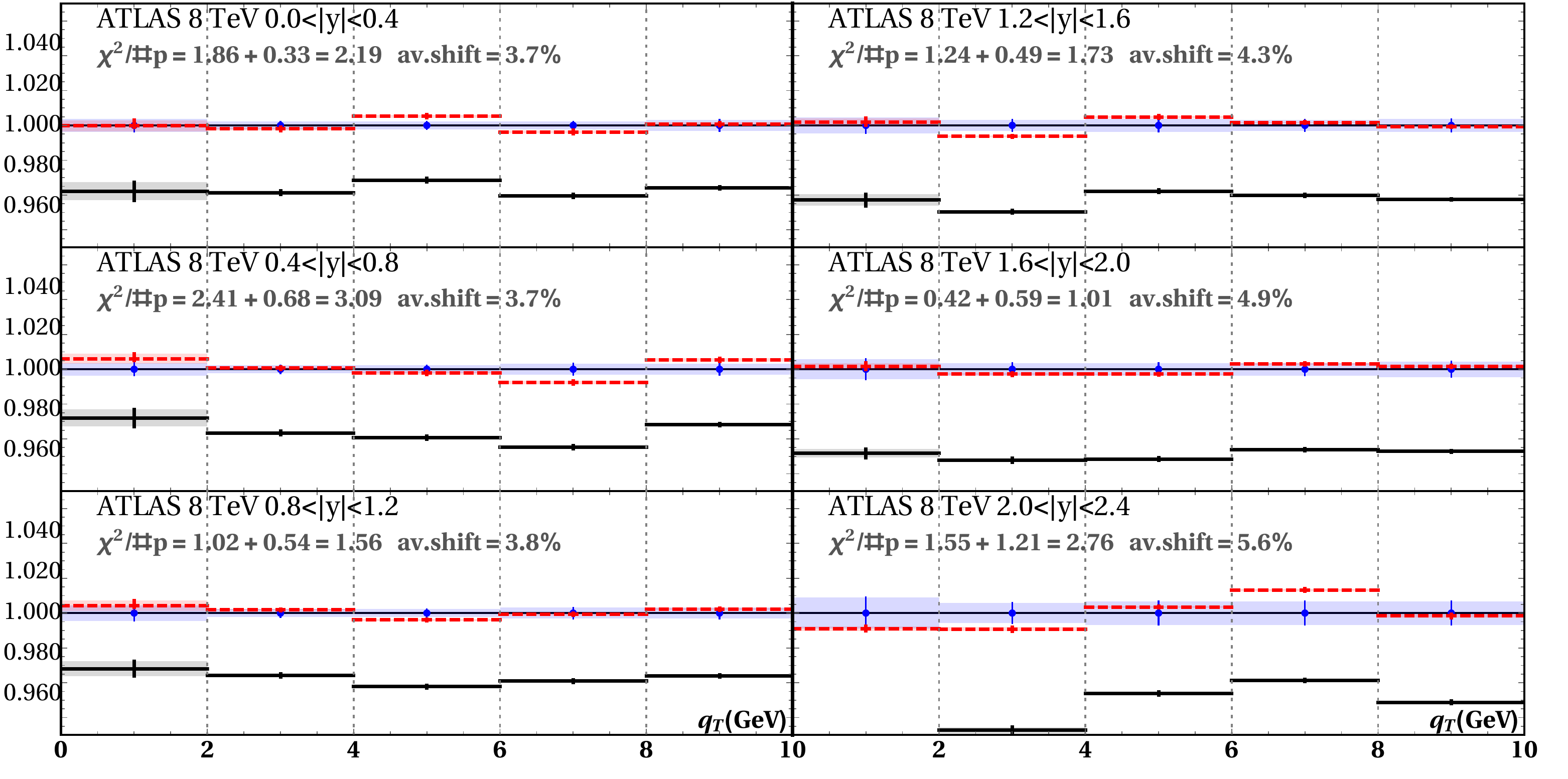}
\caption{\label{fig:atlas+}Ratio of theoretical and experimental points as a function of the binned di-lepton transverse momentum for the measured at ATLAS in the range $66<Q<116$ GeV. Black lines corresponds to the values $t_i$ predicted by the theory, whereas red dashed lines corresponds to $\bar t_i$ (\ref{eq:shiftedpreds}). The experimental points (blue dots) are surrounded by a box describing their error. For this data set, the correlated systematic uncertainty is mainly given by luminocity uncertainty is $\sim 2.8\%$ \cite{Aad:2015auj}.}
\end{center}
\end{figure}

In this appendix we present the decomposition of the $\chi^2$-value to the uncorrelated and penalty parts with the help of the so-called ``nuisance parameters''. This representation is helpful for visualization of the effect of systematic uncertainties, and allows to compute the \textit{systematic shifts}. Our presentation follows refs.\cite{Ball:2008by,Ball:2012wy}.

In order to quantify the effect of systematic uncertainties, we write the $\chi^2$ in terms of the so-called ``nuisance parameters'' $\lambda_\alpha$. It is possible
to show \cite{Ball:2012wy} that the definition of the $\chi^2$ in eq.~(\ref{eq:chi2cov})
is equivalent to
\begin{equation}\label{eq:chi2nuis}
\chi^2 = \sum_{i=1}^n\frac{1}{s_i^2}\left(m_i -t_i
  -\sum_{\alpha=1}^k\lambda_\alpha \sigma_{i,\rm corr}^{(\alpha)} \right)^2 + \sum_{\alpha=1}^k\lambda_\alpha^2\,,
\end{equation}
where $s_i^2=\sigma_{i,\rm stat}^2 +\sigma_{i,\rm unc}^2$. The optimal value of the nuisance parameters can then be determined by
minimizing the $\chi^2$ with respect to them imposing that
\begin{equation}
\frac{\partial \chi^2}{\partial \lambda_\beta} = 0\,.
\end{equation}
This yields the system
\begin{equation}\label{eq:nuissys}
  \sum_{\beta=1}^kA_{\alpha\beta}\lambda_\beta =\rho_\alpha\,,
\end{equation}
with:
\begin{equation}\label{eq:sysing}
A_{\alpha\beta}= \delta_{\alpha\beta}+\sum_{i=1}^n\frac{\sigma_{i,\rm corr}^{(\alpha)}\sigma_{i,\rm corr}^{(\beta)}}{s_i^2}\quad\mbox{and}\quad \rho_\alpha=\sum_{i=1}^n\frac{m_i-t_i}{s_i^2}\sigma_{i,\rm corr}^{(\alpha)}\,,
\end{equation}
that determines the values of $\lambda_\beta$. The quantity
\begin{equation}\label{eq:sysshiftdef}
d_i =\sum_{\alpha=1}^k\lambda_\alpha \sigma_{i,\rm corr}^{(\alpha)}
\end{equation}
in eq.~(\ref{eq:chi2nuis}) can be interpreted as a shift caused by the
correlated systematic uncertainties. As a matter of fact, defining the shifted predictions
as
\begin{equation}\label{eq:shiftedpreds}
\overline{t}_i =t_i+d_i\,,
\end{equation}
the $\chi^2$ reads
\begin{equation}\label{eq:chi2nuisshift}
  \chi^2 = \sum_{i=1}^n\left(\frac{m_i -\overline{t}_i}{s_i}\right)^2 + \sum_{\alpha=1}^k\lambda_\alpha^2=\chi_D^2+\chi_\lambda^2\,.
\end{equation}
Therefore, up to a penalty term $\chi_\lambda^2$ given by the sum of the square of the nuisance parameters, the $\chi^2$ takes the form of the uncorrelated definition $\chi_D^2$, \textit{i.e.} with diagonal covariance matrix. 

In order to achieve a visual assessment of the agreement between data and theory, it appears natural to compare the central experimental values $m_i$ to the shifted theoretical predictions $\overline{t}_i$ in units of the uncorrelated uncertainty $s_i$. The example of comparison of shifted/unshifted data is given in fig.~\ref{fig:atlas+}.

\bibliographystyle{JHEP}
\bibliography{TMD_ref}
\end{document}